\documentclass[aps,prd,10pt,a4paper,superscriptaddress,preprintnumbers,showpacs,notitlepage,eqsecnum,nofootinbib]{revtex4-1}

\usepackage[english]{babel}
\usepackage{graphicx}
\usepackage{amsmath}
\usepackage{xcolor}

\usepackage{eufrak}
\usepackage{caption}
\usepackage{subcaption}
\usepackage{hyperref}

\def\square{\mathchoice\sqr54\sqr54\sqr{2.1}3\sqr{1.5}3}
\def\sqr#1#2{{\vcenter{\vbox{\hrule height.#2pt\hbox{\vrule
width.#2pt height#1pt \kern#1pt\vrule width.#2pt}\hrule height.#2pt}}}}
\def\square{\mathchoice\sqr54\sqr54\sqr{2.1}3\sqr{1.5}3}

\makeatother

\begin{document}

\title{Chern-Weil theorem, Lovelock Lagrangians in critical dimensions\\
and boundary terms in gravity actions\\}

\author{Nathalie Deruelle$^1$, Nelson Merino$^1$ and Rodrigo Olea$^2$\\
$^1$ \it{APC, CNRS-Universit\'e Paris 7,\\
75205 Paris CEDEX 13, France\\
$^2$Departamento de Ciencias F\'{\i}sicas, Universidad
Andres Bello,\\
Sazi\'e 2212, Piso 7, Santiago, Chile}\\}

\begin{abstract}

In this paper we show how to translate into tensorial language the Chern-Weil theorem for the Lorentz symmetry, which equates the difference of the Euler densities of two manifolds to the exterior derivative of a transgression form. 
For doing so we need to introduce an auxiliary, \textit{hybrid}, manifold whose geometry we construct explicitely.
This allows us to find the vector density, constructed out of spacetime quantities only, whose divergence is the exterior derivative of the transgression form.
As a consequence we can show how the Einstein-Hilbert, Gauss-Bonnet and, in general, the Euler scalar densities can be written as the divergences of genuine vector densities in the critical dimensions $D=2,4$, etc.
As Lovelock gravity is a dimensional continuation of Euler densities, these results are of relevance for Gauss-Bonnet and, in general, Lovelock gravity.
Indeed, these vectors which can be called generalized Katz vectors ensure, in particular, a well-posed Dirichlet variational principle.
\end{abstract}

\maketitle

\section{Introduction}

It is well known that the Einstein tensor is identically zero in two dimensions and that the Gauss-Bonnet tensor is identically zero in four dimensions. 
The easiest way to show this fact in tensorial language is to write these tensors \textit{\`a la} Lovelock \cite{Lovelock} using the generalized Kronecker symbol (see also \cite{DeruelleMadore}). 
These tensors being, up to specific divergences, the variational derivatives of the Einstein-Hilbert (EH) or Gauss-Bonnet (GB) Lagrangians, a number of authors 
\cite{Yale:2010jy}, see also \cite{Cherubini:2003nj,Sotiriou:2014pfa,Santillan:2017nik,Chakraborty:2017zep},
have stated that the Lagrangians themselves could be written (in the critical dimensions two or four) as divergences of some objects, since the variational derivative of a divergence is identically zero. 

Now, since the EH and the GB Lagrangians are scalar densities containing second derivatives of the metric at most, they must be divergences of vector densities containing at most first derivatives of the metric. However, it is impossible to build a vector density out of the metric and its derivatives alone. Therefore, another ingredient must be added.
For example, in his proof \cite{Horndeski} that the Lovelock scalar densities are indeed the divergences of true to gods vector densities $V^{\mu}$, Horndeski had to introduce an arbitrary non-null contravariant vector $U^{\mu}$. 

One can also follow the formalism of Myers \cite{Myers:1987yn} to show that the Einstein and Gauss-Bonnet tensors are identically zero in the critical dimensions by relating the corresponding EH and GB actions directly to surface terms, without trying first to write them
as the divergences of vector densities. However, it turns out that Myers surface terms \textit{are} in fact the radial components of vector densities (something which is not guaranteed \textit{a priori} for any boundary term). Indeed, as we show explicitely in Appendix \ref{Myers_from_Horndeski}, 
the radial components of Horndeski's $V^{\mu}$ reproduce Myers' boundary terms in the critical dimensions, when the extra vector $U^{\mu}$ is chosen to be the normal to the boundary.

Now, whereas Horndeski's proof is purely tensorial and introduces explicitely an extra vector, Myers uses the vielbein language where the invariance under diffeomorphisms and the Lorentz symmetry are restricted to the boundary, a fact which, as we will see below, hides the necessity of introducing an extra structure. 

Our approach to show that the Lovelock scalar densities can be written as the divergences of explicit vector densities in the critical dimensions will rely on the translation of the Chern-Weil (CW) theorem (see, e.g., \cite{ChernWeil}) for the Lorentz symmetry, which is at the heart of Myers' proof, into fully covariant spacetime tensorial language. 
The CW theorem states that in $D=2p$ dimensions the difference of the Euler densities of two manifolds is equal to the exterior derivative of a $2p-1$-form, which is called a \textit{transgression} form (TF).
Since this theorem involves two different manifolds, the needed extra structure, instead of the extra vector introduced by Horndeski, will be one of the two manifolds, that we will refer to as the \textit{background}.

This translation is interesting for a number of reasons. 
First, it confirms that relating the Lorentz gauge invariance of transgressions forms and the invariance under general diffeomorphisms of boundary terms in gravity theories requires the introduction of an additional structure. 
Second, the divergences of the vector densities we shall construct, which can rightly be called \textit{generalized Katz vector densities} \cite{Katz:1985}, when added to the \textit{dimensionally continued} Lovelock actions, guarantee that their variations with respect to the metric obey Dirichlet boundary conditions.
These Katz vectors also ensure, with a proper choice of the background manifold, that the actions are finite on shell as well as the corresponding Noether charges. Indeed, it was shown in Ref.\cite{Deruelle:2017xel} (see also \cite{DKO}), for the EGB gravity case, that adding the divergence of the generalized Katz vector density to the action provides simultaneously the correct conserved charges together with a well-defined variational principle.
However, not much detail was given there about the geometrical meaning of its construction. Thus, the present work is also intended to fill this gap.

More precisely, we will show that the generalized Katz vector densities, that we shall construct with 
geometrical objects associated with two manifolds $\mathcal{M}$ and $\mathcal{\bar{M}}$, are
directly related with a transgression form constructed with the spin
connections associated with $\mathcal{M}$ and an auxiliary, \textit{hybrid}, manifold $\mathcal{\breve{M}}$, whose
geometry we shall completely characterize. 

As a consequence, we will show that if the background is chosen in such a way that the Euler density of the associated hybrid manifold vanishes, then the Einstein-Hilbert, Gauss-Bonnet and, in general, the Lovelock Lagrangians reduce to the divergence of a vector density constructed with spacetime tensors in the critical dimensions $D=2,4$, etc. 
Moreover, using Gauss coordinates for a radial foliation, the radial component of this vector reproduces Myers' boundary terms. 
This shows explicitely that, to relate Myers' terms with the divergence of vector densities constructed with spacetime quantities, an extra structure is indeed required.

This article is organized as follows. In Section \ref{Preli} we give the main
ingredients we will use, namely a brief review about the vielbein formalism
and the Chern-Weil theorem. In Section \ref{S_Issues} we explain why in
general it is not possible to make a full translation of a Lorentz transgression
form to tensorial language and analyze the differences between Lorentz and
spacetime tensors with respect to two different manifolds. Then, in Section
\ref{S_A_tens_version} we introduce the hybrid manifold that allows us to
obtain the tensorial version of the Chern-Weil theorem. Finally, Section \ref{Final} contains some further comments. \\ \\

\section{Preliminaries\label{Preli}}

The text-book material presented in this preliminary section is due to fix our conventions and notations.

\subsection{Vielbein formalism: a recap}

The vielbein $e^{A}=e_{\mu}^{A}dx^{\mu}$ and spin connection $\omega_{\ B}%
^{A}=\omega_{\mu B}^{A}dx^{\mu}$, where $x^{\mu}=t,r,\phi_{1},\ldots\phi
_{D-2}$ are spacetime coordinates and $A,B=0,\ldots,D-1$ are Lorentz indices,
are one-forms allowing to describe the geometry of a $D$-dimensional manifold
$\mathcal{M}_{D}$ in a way similar to what is done in the tensorial language by means of
the metric and the affine connection $\left(  g_{\mu\nu},\Gamma_{\mu\nu
}^{\lambda}\right)  $. The main difference is that the vielbein formulation
makes explicit reference to the local Lorentz symmetry as an internal gauge
symmetry. Denoting by $e_{A}^{\mu}$ the inverse matrix of the vielbein
components $e_{\mu}^{A}$, such that $e_{A}^{\mu}e_{\nu}^{A}=\delta_{\nu}^{\mu}$
and $e_{A}^{\mu}e_{\mu}^{B}=\delta_{A}^{B}\,$, the relation between both
languages is given by,%
\begin{align}
\eta_{AB}  &  =e_{A}^{\mu}e_{B}^{\nu}g_{\mu\nu}\,,\label{df_1}\\
\omega_{\mu B}^{A}  &  =e_{\alpha}^{A}e_{B}^{\gamma}\Gamma_{\mu\gamma}%
^{\alpha}+e_{\alpha}^{A}\partial_{\mu}e_{B}^{\alpha}\,. \label{df_1b}%
\end{align}
The first relation (\ref{df_1}) states that in each point of $\mathcal{M}_{D}$ it is possible to find an invertible coordinate transformation $x^{\mu}=x^{\mu}(y^{A})$ such that the Jacobian matrix $e_{A}^{\mu}=\partial x^{\mu}/\partial
y^{A}$ brings $g_{\mu\nu}$ to the Minkowski metric $\eta_{AB}$. Thus, the
vielbein components $e_{\mu}^{A}$ is the Jacobian of the inverse
transformation $e_{\mu}^{A}=\partial y^{A}/\partial x^{\mu}\,$. By
construction, the vielbein $e^{A}$ and spin connection $\omega_{\ B}^{A}$ are
invariant under coordinate transformations $x^{\prime\mu}=x^{\prime\mu}%
(x^{\nu})$, while under a local Lorentz transformation $y^{\prime A}=\Lambda_{\ B}^{A}y^{B}$ (with $\Lambda^{T}\eta\Lambda=\eta$ and $\eta$ being
the Lorentz metric) they transform respectively as%
\begin{align}
e^{\prime A}  &  =\Lambda_{\ B}^{A}e^{B}\,,\label{transf_e}\\
\omega_{\ \ B}^{\prime A}  &  =\Lambda_{\ C}^{A}\Lambda_{B}^{\ D}\omega
_{\ D}^{C}+\Lambda_{\ C}^{A}d\Lambda_{B}^{\ C}\,, \label{transf_w}%
\end{align}
where $\Lambda_{B}^{\ A}$ denotes the inverse of $\Lambda_{\ B}^{A}$ and $d$ is the exterior derivative. In particular, given a metric tensor $g_{\mu\nu}$, the vielbeins can be
determined up to a Lorentz transformation and therefore $e_{\mu}^{A}$ carries
the same number of independent components as $g_{\mu\nu}$.

The second relation (\ref{df_1b}), also known as the \textit{tetrad
postulate}, implies that the curvature and torsion two-forms defined as
\begin{align}
\Omega_{\ B}^{A}  &  \equiv d\omega_{\ B}^{A}+\omega_{\ C}^{A}\omega_{\ B}%
^{C}=\frac{1}{2}\Omega_{\ B\mu\nu}^{A}dx^{\mu}dx^{\nu}\,,\label{df_2}\\
T^{A}  &  \equiv De^{A}=\frac{1}{2}T_{\mu\nu}^{A}dx^{\mu}dx^{\nu}\,,\text{
\ \ with \ \ }De^{A}=de^{A}+\omega_{\ B}^{A}e^{B}\,, \label{df_2b}%
\end{align}
where $D$ defines the
Lorentz covariant derivative, are related with the Riemann and torsion tensors
$R_{\ \beta\mu\nu}^{\alpha}=\partial_{\mu}\Gamma_{\beta\nu}^{\alpha}-\ldots$
and $T_{\mu\nu}^{\lambda}=\Gamma_{\mu\nu}^{\lambda}-\Gamma_{\nu\mu}^{\lambda}$ by
\begin{equation}
\Omega_{\ B\mu\nu}^{A}=e_{\alpha}^{A}e_{B}^{\beta}R_{\ \beta\mu\nu}^{\alpha
}\ \ \ \text{and\ \ \ }T_{\mu\nu}^{A}=e_{\lambda}^{A}T_{\mu\nu}^{\lambda}\,.
\label{df_2c}%
\end{equation}
With this notation the Ricci scalar is given by $R=g^{\mu\nu}R_{\mu\nu}$ with
$R_{\mu\nu}=R_{\ \mu\alpha\nu}^{\alpha}$ being the Ricci tensor. Remark that, for shortness, we omit the wedge product between differential forms.

The manifold $\mathcal{M}_{D}$ is pseudo-Riemannian if it
satisfies the metricity condition $\nabla_{\lambda}g_{\mu\nu}=0$
(here $\nabla$ stands for covariant derivative with respect to $\Gamma$) and
the torsionless condition $T_{\mu\nu}^{\lambda}=0$. The only connection
satisfying simutaneously these conditions is the Christoffel connection, which
is completely determined by the metric and its derivatives, $\Gamma_{\mu
\gamma}^{\alpha}=\Gamma_{\mu\gamma}^{\alpha}\left(  g,\partial g\right)  $.
Similarly, in the vielbein formalism, a pseudo-Riemannian geometry is
characterized by the conditions $D\eta_{AB}=0$ and $T^{A}=0$. The first one is
equivalent to assume that the spin connection is antisymmetric (i.e.,
$\omega^{AB}=-\omega^{BA}$ with $\omega^{AB}=\eta^{BC}\omega_{\ C}^{A}$) and together with the torsionless condition
one is able to solve the spin connection in terms of the vielbein
and its derivatives, $\omega_{\mu B}^{A}=\omega_{\mu B}^{A}\left(  e,\partial
e\right)  $, the explicit expression of which will not be needed here.

Finally, the Levi-Civita symbols $\varepsilon_{\mu_{1}\ldots\mu_{D}}$ and
$\varepsilon_{A_{1}\ldots A_{D}}$ together with $\varepsilon^{\mu_{1}\ldots
\mu_{D}}$ and $\varepsilon^{A_{1}\ldots A_{D}}$ are such that $\varepsilon
^{\mu_{1}\ldots\mu_{D}}=-\varepsilon_{\mu_{1}\ldots\mu_{D}}$ and
$\varepsilon^{A_{1}\ldots A_{D}}=-\varepsilon_{A_{1}\ldots A_{D}}$ with the
convention $\varepsilon_{tr\theta_{1}\ldots\theta_{D-2}}=\varepsilon
_{012\ldots D}=1$. It is easy to show that under a coordinate tranformations
$\epsilon_{\mu_{1}\ldots\mu_{D}}$ and $\epsilon^{\mu_{1}\ldots\mu_{D}}$
transform respectively as tensorial densities of weight $1$ and $-1$. As for
$\varepsilon_{A_{1}\ldots A_{D}}$ and $\varepsilon^{A_{1}\ldots A_{D}}$ they
transform as tensors under local Lorentz transformations. Moreover, both are
related by%
\begin{equation}
\varepsilon_{A_{1}\ldots A_{D}}e_{\mu_{1}}^{A_{1}}\cdots e_{\mu_{D}}^{A_{D}%
}=\sqrt{-g}\,\varepsilon_{\mu_{1}\ldots\mu_{D}}\,, \label{df_3}%
\end{equation}
where $e_{\mu}^{A}=\partial y^{A}/\partial x^{\mu}\,$.

\subsection{Chern-Weil theorem\label{S_CW}}

The Chern-Weil theorem (see, e.g., \cite{ChernWeil}) was developed in quest for a proof of the
generalized Gauss-Bonnet theorem. It is regarded a milestone
towards a complete theory of characteristic classes which relates and unifies
concepts in algebraic topology and differential geometry. It is formulated in
terms of fiber bundle structures, a powerful tool that allows to
build a gauge theory over a smooth manifold. Its basic ingredients are a Lie algebra with generators $T_{M}$, a Lie valued gauge
connection one-form $A$ 
%before it was written: $A=A_{\mu}^{M}T_{M}dx^{\mu}$
and its corresponding field strength $F=dA+A\wedge A$. It is easy to show that 
$\left\langle F^{p}\right\rangle $, where $\left\langle \ \right\rangle $ stands for the symmetrized trace of the generators, is invariant under gauge transformations in $2p$ dimensions and thus, it is a topological term. 
The Chern-Weil theorem states that, given two connections $A$ and $\bar{A}$, the topological terms constructed
with their corresponding curvatures are closed forms and that 
the difference $\left\langle  F^{p}\right\rangle - \left\langle \bar{F}^{p}\right\rangle$ is an exact form, i.e., is the exterior derivative of an odd-form which is known as
\textit{transgression} form (see Appendix \ref{App_Tran} for its general expression).  
In particular, a Chern-Simons form is recovered from a transgression form by setting the second connection to zero.

In the case where the symmetry is described by the Lorentz algebra, the \textit{Euler} topological term for a $2p$-dimensional
pseudo-Riemannian manifold $\mathcal{M}_{2p}$ (with $p$ an integer) is defined in the vielbein formalism
as
\begin{equation}
\mathcal{E}_{2p}\left(  \Omega\right)  \equiv\varepsilon_{A_{1}\ldots A_{2p}%
}\Omega^{A_{1}A_{2}}\cdots\Omega^{A_{2p-1}A_{2p}}\,, \label{CW_1}%
\end{equation}
where $\Omega^{AB}=\eta^{AC}\Omega_{\ C}^{A}$. This quantity is, by
construction, a $2p$-form invariant under local Lorentz transformations. It is a topological term because, as stated by the Gauss-Bonnet
theorem, its integral over a compact manifold is related with the Euler
characteristic $\chi\left(  \mathcal{M}_{2p}\right)  $ which describes its topology. Further details can be found, e.g., in Ref.
\cite{Miskovic:2007mg}.

Consider now a second pseudo-Riemannian manifold $\mathcal{\bar{M}}_{2p}$ with
Lorentz connection $\bar{\omega}_{\ B}^{A}$, curvature $\bar{\Omega}_{\ B}%
^{A}=d\bar{\omega}_{\ B}^{A}+\bar{\omega}_{\ C}^{A}\bar{\omega}_{\ B}^{C}$.
Using that by definition the Lorentz tensors $\eta_{AB}$ and $\varepsilon
_{A_{1}\ldots A_{2p}}$ are the same for both manifolds (because the Minkowski
tangent space is the same for each point of each manifold), we can define
$\bar{\omega}^{AB}=\eta^{BC}\bar{\omega}_{\ C}^{A}$ and $\bar{\Omega}%
^{AB}=\eta^{BC}\bar{\Omega}_{\ C}^{A}$ so that the Euler term in
$\mathcal{\bar{M}}_{2p}$ is given by%
\begin{equation}
\mathcal{E}_{2p}(\bar{\Omega})=\varepsilon_{A_{1}\ldots A_{2p}}\bar{\Omega
}^{A_{1}A_{2}}\cdots\bar{\Omega}^{A_{2p-1}A_{2p}}\,. \label{CW_1b}%
\end{equation}
Now, the \textit{Chern-Weil (CW) theorem} for the Lorentz symmetry establishes that
the difference between the two topological terms (\ref{CW_1}) and
(\ref{CW_1b}) is an exact form, i.e., the exterior derivative of a $\left(
2p-1\right)  $-form $\mathcal{T}^{\left(  2p-1\right)  }$, called
\textit{transgression} form, which is completely determined by the connections
$\omega$ and $\bar{\omega}$:%

\begin{equation}
\mathcal{E}_{2p}\left(  \Omega\right)  -\mathcal{E}_{2p}(\bar{\Omega
})=d\mathcal{T}^{\left(  2p-1\right)  }\,. \label{CW_2}%
\end{equation}

For example, for $p=1$ the anstisymmetric property of the spin connections
$\omega$ and $\bar{\omega}$ leads $\Omega^{AB}=d\omega^{AB}$ and $\bar{\Omega
}^{AB}=d\bar{\omega}^{AB}$ and thus, the difference of the two Euler
terms $\mathcal{E}_{2}\left(  \Omega\right)  =\varepsilon_{AB}\Omega
^{AB}$ and $\mathcal{E}_{2}\left(  \bar{\Omega}\right)  =\varepsilon_{AB}%
\bar{\Omega}^{AB}$ is simply given by%

\begin{equation}
\mathcal{E}_{2}\left(  \Omega\right)  -\mathcal{E}_{2}(\bar{\Omega
})=d[\mathcal{T}^{\left(  1\right)  }(\bar{\theta})]\,,\text{ \ \ with
\ \ }\mathcal{T}^{\left(  1\right)  }(\bar{\theta})=\varepsilon_{AB}%
\bar{\theta}^{AB}\text{ \ \ and \ \ }\bar{\theta}^{AB}\equiv\omega^{AB}%
-\bar{\omega}^{AB}\,. \label{cw2_gen_1}%
\end{equation}
This is the simplest realization of the Chern-Weil theorem for the Lorentz symmetry.

As reviewed in the Appendix \ref{App_Tran}, for higher values of $p\,$the
transgression form is given by%
\begin{equation}
\mathcal{T}^{\left(  2p-1\right)  }(\bar{\theta},\Omega,\bar{\Omega}%
)=p\int_{0}^{1}dt\,\varepsilon_{A_{1}\ldots A_{2p}}\bar{\theta}^{A_{1}A_{2}%
}\Omega_{\left(  t\right)  }^{A_{3}A_{4}}\cdots\,\Omega_{\left(  t\right)
}^{A_{2p-1}A_{2p}}\,, \label{Tran}%
\end{equation}
where $\Omega_{\left(  t\right)  }^{AB}=d\omega_{\left(  t\right)  }%
^{AB}+\omega_{\left(  t\right)  C}^{A}\omega_{\left(  t\right)  }^{CB}$ and
$\omega_{\left(  t\right)  }^{AB}=\bar{\omega}^{AB}+t\bar{\theta}^{AB}$ is a
connection which interpolates between $\bar{\omega}^{AB}$ and $\omega^{AB}$
for $t\in\left[  0,1\right]  $. In Appendix \ref{App_Tran} it is also shown
that the interpolating curvature has the following alternative expressions,%
\begin{align}
\Omega_{\left(  t\right)  }^{AB}  &  =t\Omega^{AB}+\left(  1-t\right)
\bar{\Omega}^{AB}-t\left(  1-t\right)  \bar{\theta}_{\ C}^{A}\bar{\theta}%
^{CB}\label{int_curv_a}\\
&  =\bar{\Omega}^{AB}+t\bar{D}\bar{\theta}^{AB}+t^{2}\bar{\theta}_{\ C}%
^{A}\bar{\theta}^{CB}\label{int_curv_b}\\
&  =\Omega^{AB}+\left(  t-1\right)  D\bar{\theta}^{AB}+\left(  t-1\right)
^{2}\bar{\theta}_{\ C}^{A}\bar{\theta}^{CB} \label{int_curv_c}\,,
\end{align}
where in the last two lines $\bar{D}$ and $D$, which represent respectively the
covariant derivatives with respect to the connections $\bar{\omega}$ and $\omega$,
are related as $\bar{D}\bar{\theta}^{AB}=D\bar{\theta}^{AB}%
-2\bar{\theta}_{\ C}^{A}\bar{\theta}^{CB}$.\\

Transgression forms (TFs) have proved to be useful to deal with a number of
different physical situations. Originally used to treat the general
problem of anomalies in field theory \cite{Manes:1985df,Zumino:1983ew,AlvarezGaume:1984dr}, more
recent applications range from the study of anomalies in hydrodynamics in the
context of gauge/gravity duality \cite{Haehl:2013hoa} to holographic models of baryons \cite{Lau:2016dxk}.

In the context of gravity, the use of TFs is possible whenever the Lie algebra
accounts for the symmetries of the Lagrangian (e.g., Lorentz, (anti-)de
Sitter, etc.). In particular, TFs have been used in a dimensionally continued
version to define a well-posed variational principle in different gravity
theories: The Gibbons-Hawking-York (GHY) boundary term
\cite{York:1972sj,Gibbons:1976ue}, and its generalization by Myers
\cite{Myers:1987yn} to the case of Lovelock theories \cite{Lovelock},
that defines a Dirichlet problem in Einstein / Lovelock gravity can be derived
from a TF for the Lorentz group. In that construction, the first connection is
the spin connection associated with the dynamical spacetime, while the second
one is associated with a product manifold with precise matching
conditions at the boundary \cite{Eguchi:1980jx}. 
Thus, when the symmetry is reduced to the boundary, the information about the product manifold disappears and a well-defined action principle is set without the need of an explicit background geometry (see, e.g, \cite{Miskovic:2007mg}). However, both the
action and its variation are usually divergent on shell for asymptotically (anti-de Sitter) solutions.  Other applications of TFs in gravity can be found, e.g, in Refs. \cite{Izaurieta:2006wv,Izaurieta:2006aj,Merino:2009ya,Merino:2010zz,Izaurieta:2011fr,Salgado:2014jka}.

\section{Issues about the tensorial translation\label{S_Issues}}

\subsection{The problem\label{S_Issues_A}}

As it has been reviewed in the previous section, given two pseudo-Riemannian
manifolds $\mathcal{M}_{2p}$ and $\mathcal{\bar{M}}_{2p}$ of dimension $2p$,
the Chern-Weil theorem states that:%
\begin{equation}
\mathcal{E}_{2p}\left(  \Omega\right)  -\mathcal{E}_{2p}(\bar{\Omega
})=d\mathcal{T}^{\left(  2p-1\right)  }\,, \label{CW_}%
\end{equation}
where $\mathcal{E}_{2p}\left(  \Omega\right)  $, $\mathcal{E}_{2p}(\bar
{\Omega})$ are the topological terms defined by (\ref{CW_1},\ref{CW_1b})
and where $\mathcal{T}^{\left(  2p-1\right)  }$ is the transgression form
defined by (\ref{Tran}), which depends on the connections $\omega$ and
$\bar{\omega}$ through the Lorentz covariant objects $\Omega^{AB}$,
$\bar{\Omega}^{AB}$ and $\bar{\theta}^{AB}$ (see Eqs. (\ref{int_curv_a}%
-\ref{int_curv_c})). The l.h.s. of (\ref{CW_}) can be translated rightaway into tensorial
language as%
\begin{equation}
\mathcal{E}_{2p}\left(  \Omega\right)  -\mathcal{E}_{2p}(\bar{\Omega}%
)=\frac{1}{2^{p}}\delta_{\alpha_{1}\ldots\alpha_{2p}}^{\mu_{1}\ldots\mu_{2p}%
}\left[  \sqrt{-g}\,R_{\mu_{1}\mu_{2}}^{\alpha_{1}\alpha_{2}}\cdots
R_{\mu_{2p-1}\mu_{2p}}^{\alpha_{2p-1}\alpha_{2p}}-\sqrt{-\bar{g}}\,\bar
{R}_{\mu_{1}\mu_{2}}^{\alpha_{1}\alpha_{2}}\cdots\bar{R}_{\mu_{2p-1}\mu_{2p}%
}^{\alpha_{2p-1}\alpha_{2p}}\right]  d^{2p}x\,, \label{CW_3}%
\end{equation}
i.e., it can be written as an expression depending only on spacetime tensorial
objects such as $g_{\mu\nu}\,$, $\bar{g}_{\mu\nu}\,$, $R_{\ \gamma\mu\nu}^{\alpha}$
and $\bar{R}_{\ \gamma\mu\nu}^{\alpha}\,$. The translation can be easily made using the relations
\begin{align}
\Omega^{AB}  &  =\frac{1}{2}e_{\alpha}^{A}e_{\beta}^{B}R_{\ \ \mu\nu}%
^{\alpha\beta}dx^{\mu}dx^{\nu}\,,\ \ \ \bar{\Omega}^{AB}=\frac{1}{2}\bar
{e}_{\alpha}^{A}\bar{e}^{B\beta}\bar{R}_{\ \beta\mu\nu}^{\alpha}dx^{\mu
}dx^{\nu}\,,\label{CW_4}\\
R_{\ \ \mu\nu}^{\alpha\beta}  &  =g^{\beta\gamma}R_{\ \gamma\mu\nu}^{\alpha
}\,,\ \ \ \bar{R}_{\ \ \mu\nu}^{\alpha\beta}=\bar{g}^{\beta\gamma}\bar
{R}_{\ \gamma\mu\nu}^{\alpha}\,,\nonumber
\end{align}
which hold due to the tetrad postulates $\omega_{\mu}^{AB}=e_{\alpha}%
^{A}e^{B\gamma}\Gamma_{\mu\gamma}^{\alpha}+e_{\alpha}^{A}\partial_{\mu
}e^{B\alpha}$ and $\bar{\omega}_{\mu}^{AB}=\bar{e}_{\alpha}^{A}\bar
{e}^{B\gamma}\bar{\Gamma}_{\mu\gamma}^{\alpha}+\bar{e}_{\alpha}^{A}%
\partial_{\mu}\bar{e}^{B\alpha}\,$, together with the identities%
\begin{align}
\varepsilon_{A_{1}\ldots A_{2p}}e_{\mu_{1}}^{A_{1}}\cdots e_{\mu_{2p}}%
^{A_{2p}}  &  =\sqrt{-g}\,\varepsilon_{\mu_{1}\ldots\mu_{2p}}%
\,,\ \ \ \varepsilon_{A_{1}\ldots A_{2p}}\bar{e}_{\mu_{1}}^{A_{1}}\cdots
\bar{e}_{\mu_{2p}}^{A_{2p}}=\sqrt{-\bar{g}}\,\varepsilon_{\mu_{1}\ldots
\mu_{2p}}\,, \nonumber\\
\varepsilon_{\mu_{1}\ldots\mu_{2p}}\varepsilon^{\nu_{1}\ldots\nu_{2p}}  &
=-\delta_{\mu_{1}\ldots\mu_{2p}}^{\nu_{1}\ldots\nu_{2p}}\,,\ \ \ dx^{\mu_{1}%
}\cdots dx^{\mu_{2p}}=-\varepsilon^{\mu_{1}\ldots\mu_{2p}}d^{2p}x\,,
\label{CW_4b}%
\end{align}
with $d^{2p}x=dtdrd\phi_{1}\ldots d\phi_{2p-2}\,$\ and $\delta_{\mu_{1}%
\ldots\mu_{2p}}^{\nu_{1}\ldots\nu_{2p}}$ being the generalized Kronecker delta
defined as the determinant of the $2p\times2p$ matrix $\left(  \delta_{\mu
_{s}}^{\nu_{r}}\right)  $ with $r,s=1,\ldots,2p\,$.

On the other hand, a complete translation in terms of spacetime tensors only of the r.h.s of (\ref{CW_}) is not possible working with $\cal{M}$ and $\mathcal{\bar{M}}$ only. Indeed, the only thing that can be shown is that
\[
d[\mathcal{T}^{\left(  2p-1\right)  }(\bar{\theta},\Omega,\bar{\Omega
})]=\partial_{\mu}[v^{\mu}(\bar{\theta},\Omega,\bar{\Omega})]d^{2p}x\,,
\]
where $v^{\mu}$ is a spacetime vector density which depends on $\Gamma
,\bar{\Gamma}$ but \textit{also} on $e,\bar{e}\,$: $v^{\mu}=v^{\mu}(e,\Gamma,\bar{e}%
,\bar{\Gamma})$ and, in general, there is no way to get rid of the vielbeins.

For example, in $D=2$, the l.h.s. of Eq. (\ref{cw2_gen_1}) is given by $(\sqrt{-g}R-\sqrt{-\bar{g}}\bar{R}\,)d^{2}x\,$, while the r.h.s. is given by
$\partial_{\mu}v^{\mu}d^{2}x$ where $v^{\mu}$ is the following vector
density\footnote{The quantity $v^{\mu}$ is indeed a vector density because,
for spacetime transformations, $\bar{\theta}_{\nu}^{AB}$ is a covariant vector
while $\varepsilon^{\mu\nu}$ is a contravariant tensor density. It is also
clear that for Lorentz transformations $v^{\mu}$ behaves as an invariant
because $\varepsilon_{AB}$ is a Lorentz covariant tensor or rank $2$, while
$\bar{\theta}_{\nu}^{AB}$ is a contravariant tensor of rank $2$.}
\begin{equation}
v^{\mu}=-\varepsilon_{AB}\varepsilon^{\mu\nu} \bar{\theta}_{\nu}^{AB}=-\varepsilon_{AB}\varepsilon^{\mu\nu}\left(  \omega_{\nu}^{AB}-\bar{\omega}_{\nu}^{AB}\right)  
=-\varepsilon_{AB}\varepsilon^{\mu\nu}\left[ \left( e_{\alpha}^{A}e^{B\gamma}\Gamma_{\nu\gamma}^{\alpha}
- \bar{e}_{\alpha}^{A}\bar{e}^{B\gamma}\bar{\Gamma}_{\nu\gamma}^{\alpha} \right) +\left( e_{\alpha}^{A}\partial_{\nu}e^{B\alpha} - \bar{e}_{\alpha}^{A}\partial_{\nu}\bar{e}^{B\alpha} \right) \right]  \,, 
\label{cw2_gen_2}
\end{equation}
where $\bar{\theta}_{\nu}^{AB}=\omega_{\nu}^{AB}-\bar{\omega}_{\nu}^{AB}$ are
the components of the one-form $\bar{\theta}^{AB}$ and where we have used the tetrad
postulate in the last equality. Thus $v^{\mu}=v^{\mu}%
(e,\Gamma,\bar{e},\bar{\Gamma})$ and there is at first sight no way to
completely translate $d[\mathcal{T}^{\left(  1\right)  }(\bar{\theta})]$ to
tensorial language\footnote{However, as shown in Appendix \ref{Conf_spaces}, a
version of the Chern-Weil theorem which is free of vielbeins can be formulated
in the particular case $D=2$ using the fact that in this dimension all the metrics
are conformally equivalent.}. One may orient the corresponding
vielbeins such that the $e^{-1}\partial e$ (inhomogeneous) terms vanish \cite{Charmousis:2005ey}. However, even in that case, the
resulting expression cannot be written in terms of $\Delta_{\mu\nu}^{\alpha}$,
defined as%
\begin{equation}
\Delta_{\mu\nu}^{\alpha}\equiv\Gamma_{\mu\nu}^{\alpha}-\bar{\Gamma}_{\mu\nu
}^{\alpha}\,.
\label{first_Delta}
\end{equation}
%which, as we will see in the next section, is the only tensor that can be constructed in terms of the Christoffel symbol (and not its derivatives).
Indeed the factor $(ee\Gamma-\bar e \bar e \bar\Gamma)$, which then captures the structure of $v^{\mu}$, cannot be transformed into a expression of the type $ee (\Gamma-\bar \Gamma)$ which would transform it in a spacetime tensor density because, as we show in the next subsection, $e$ and $\bar e$ are different and hence cannot be related by a local Lorentz rotation.

\subsection{Lorentz versus spacetime tensors \label{S_L_vs_S}}

Consider a pair of $D$-dimensional pseudo-Riemannian manifolds $\left\{
\,\mathcal{M},\mathcal{\bar{M}}\,\right\}  $ endowed respectively with the
metrics $\left\{  \,g_{\mu\nu},\bar{g}_{\mu\nu}\,\right\}  $ and where
$\left\{  \,\Gamma_{\mu\nu}^{\alpha},\bar{\Gamma}_{\mu\nu}^{\alpha}\,\right\}
$ are the corresponding Christoffel symbols. Also choose a mapping $\sigma$
between these manifolds allowing us to use the same coordinates $x^{\mu}$ for
each point $P\in\mathcal{M}$ and $\bar{P}=\sigma\left(  P\right)
\in\mathcal{\bar{M}}\,$. This choice is always possible and, as a consequence,
a coordinate transformation $x^{\prime\mu}=x^{\prime\mu}\left(  x^{\nu
}\right)  $ in $\mathcal{M}$ induces the same coordinate transformation in
$\mathcal{\bar{M}}$. This means that if $P_{\nu_{1}\nu_{2}\ldots}^{\mu_{1}%
\mu_{2}\ldots}$ and $\bar{P}_{\nu_{1}\nu_{2}\ldots}^{\mu_{1}\mu_{2}\ldots}$
are tensors defined respectively on $\mathcal{M}$ and $\mathcal{\bar{M}}$, then their linear combinations $aP_{\nu_{1}\nu_{2}\ldots}^{\mu_{1}\mu_{2}\ldots
}+b\bar{P}_{\nu_{1}\nu_{2}\ldots}^{\mu_{1}\mu_{2}\ldots}$ are also true
tensors, because the Jacobian matrices are the same,
\begin{equation}
a\,P^{\prime}\,_{\nu_{1}\ldots\nu_{q}}^{\mu_{1}\ldots\mu_{p}}\left(
x^{\prime}\right)  +b\,\bar{P}^{\prime}\,_{\nu_{1}\ldots\nu_{q}}^{\mu
_{1}\ldots\mu_{p}}\left(  x^{\prime}\right)  =\frac{\partial x^{\prime\mu_{1}%
}}{\partial x^{\alpha_{1}}}\cdots\frac{\partial x^{\prime\mu_{p}}}{\partial
x^{\alpha_{p}}}\frac{\partial x^{\prime\beta_{1}}}{\partial x^{\nu_{1}}}%
\cdots\frac{\partial x^{\prime\beta_{q}}}{\partial x^{\nu_{q}}}\left(
a\,P\,_{\beta_{1}\ldots\beta_{q}}^{\alpha_{1}\ldots\alpha_{p}}\left(
x\right)  +b\,\bar{P}\,_{\beta_{1}\ldots\beta_{q}}^{\alpha_{1}\ldots\alpha
_{p}}\left(  x\right)  \right)  \,. \label{L_vs_S_1}%
\end{equation}
Thus, we can deal with linear combinations of tensors, defined on different manifolds, that are simultaneously spacetime tensors on both manifolds. 
Among those tensors we have, e.g.,
\begin{align}
Y_{\mu\nu}  &  =ag_{\mu\nu}+b\bar{g}_{\mu\nu}\,,\ \ \ a,b\text{ being
arbitrary constants,}\nonumber\\
\Delta_{\mu\nu}^{\alpha}  &  =\Gamma_{\mu\nu}^{\alpha}-\bar{\Gamma}_{\mu\nu
}^{\alpha}\,, \label{L_vs_S_2}%
\end{align}
as well as derived quantities, such as 
$\Delta_{\mu}^{\alpha\beta}=g^{\beta\nu}\Delta_{\mu\nu}^{\alpha}$ and $\nabla_{\mu}\Delta_{\nu}^{\alpha\beta}$.

A similar analysis can be done in the vielbein formulation, where the analogs of (\ref{L_vs_S_2}) are\footnote{Indeed, using (\ref{transf_e}) and (\ref{transf_w}) is easy to
show that under a Lorentz transformation $E^{\prime A}=\Lambda_{\ B}^{A}E^{B}$
and $\bar{\theta}_{\ \ B}^{\prime A}=\Lambda_{\ C}^{A}\Lambda_{B}^{\ D}%
\,\bar{\theta}_{\ D}^{C}$.}%
\begin{align}
E^{A}  &  =ae^{A}+b\bar{e}^{A}\,,\ \ \ a,b\text{ being arbitrary
constants,}\nonumber\\
\bar{\theta}_{\ B}^{A}  &  =\omega_{\ B}^{A}-\bar{\omega}_{\ B}^{A}\,.
\label{L_vs_S_3}%
\end{align}
We recall that by definition the Lorentz tensors $\eta_{AB}$ and
$\varepsilon_{A_{1}\ldots A_{D}}$ are the same for both manifolds and thus we
have, for example,%
\begin{equation}
\omega^{AB}=\eta^{BC}\omega_{\ C}^{A}\,,\ \ \ \Omega^{AB}=\eta^{BC}%
\Omega_{\ C}^{A}\,,\ \ \ \bar{\omega}^{AB}=\eta^{BC}\bar{\omega}_{\ C}%
^{A}\,,\ \ \ \bar{\Omega}^{AB}=\eta^{BC}\bar{\Omega}_{\ C}^{A}\,.
\label{L_vs_S_3b}%
\end{equation}
In particular, we recognize that the one-form $\bar{\theta}^{AB}=\eta^{BC}%
\bar{\theta}_{\ C}^{A}$ is the one appearing as a fundamental object in the
definition of the transgression form $\mathcal{T}^{\left(  2p-1\right)
}\left(  \bar{\theta},\Omega,\bar{\Omega}\right)  $ given in (\ref{Tran}).

Now, the question is to determine in which cases a local Lorentz-invariant
quantity, constructed with the Lorentz-covariant objects%
\begin{equation}
\{\,\eta_{AB},\varepsilon_{A_{1}\ldots A_{D}},e^{A},\bar{e}^{A},E^{A}%
,\Omega^{AB},\bar{\Omega}^{AB},\bar{\theta}_{\ B}^{A}\,\}\,, \label{L_objects}%
\end{equation}
can be written in terms of the spacetime quantities%
\begin{equation}
\{\,\varepsilon_{\mu_{1}\ldots\mu_{D}},g_{\mu\nu},\bar{g}_{\mu\nu},Y_{\mu\nu
},\Delta_{\mu\nu}^{\alpha}\,\}\,, \label{S_objects}%
\end{equation}
knowing that the following identities hold (see Eqs. (\ref{df_1b})
and (\ref{df_3}))%
\begin{align}
\varepsilon_{A_{1}\ldots A_{D}}e_{\mu_{1}}^{A_{1}}\cdots e_{\mu_{D}}^{A_{D}}
&  =\sqrt{-g}\,\varepsilon_{\mu_{1}\ldots\mu_{D}}\,,\label{L_vs_S_4_0}\\
\varepsilon_{A_{1}\ldots A_{D}}\bar{e}_{\mu_{1}}^{A_{1}}\cdots\bar{e}_{\mu
_{D}}^{A_{D}}  &  =\sqrt{-\bar{g}}\,\varepsilon_{\mu_{1}\ldots\mu_{D}}\,,
\label{L_vs_S_4}%
\end{align}%
\begin{equation}
\omega_{\mu}^{AB}=e_{\alpha}^{A}e^{B\gamma}\Gamma_{\mu\gamma}^{\alpha
}+e_{\alpha}^{A}\partial_{\mu}e^{B\alpha}\ ,\ \ \ \bar{\omega}_{\mu}^{AB}%
=\bar{e}_{\alpha}^{A}\bar{e}^{B\gamma}\bar{\Gamma}_{\mu\gamma}^{\alpha}%
+\bar{e}_{\alpha}^{A}\partial_{\mu}\bar{e}^{B\alpha}\,, \label{L_vs_S_5}%
\end{equation}
and that, as shown in Appendix \ref{App_Rel_vielb}, the vielbeins are related
by%
\begin{equation}
e^{A}=K_{\ B}^{A}\bar{e}^{B}\,, \label{L_vs_S_5b}%
\end{equation}
where $K=K_{\ B}^{A}\left(  x\right)  $\ is not a Lorentz rotation (i.e.,
$K^{T}\eta K\neq\eta\,$) because $e$ and $\bar e$ are inequivalent as they describe different geometries. \\

The Euler terms $\mathcal{E}_{2p}\left(\Omega\right)  $ and $\mathcal{E}_{2p}\left(  \bar{\Omega}\right)$ in Section \ref{S_Issues_A} are examples where a complete translation is
possible. They depend only on $\varepsilon_{A_{1}\ldots A_{2p}}$ and the
curvatures $\Omega^{AB}$ and $\bar{\Omega}^{AB}$. Thus, the use of
(\ref{L_vs_S_4_0}-\ref{L_vs_S_5}) allows us to translate these Euler terms in
tensorial expressions free of Lorentz indices. On the other hand, in the same section, we have
found problems to translate in tensorial language the exterior derivative of
the transgression form $\mathcal{T}^{\left(  2p-1\right)  }\left(  \omega
,\bar{\omega}\right)  $ defined in (\ref{Tran}), which depends on the object (written here, for visual simplicity, in the case the vielbeins are properly oriented)
\begin{equation}
\bar{\theta}^{AB}=\eta^{BC}\bar{\theta}_{\ C}^{A}=\left[  e_{\alpha}^{A}e^{B\gamma}\Gamma_{\mu\gamma}^{\alpha}
-\bar{e}_{\alpha}^{A}\bar{e}^{B\gamma}\bar{\Gamma}_{\mu\gamma}^{\alpha}\right]
dx^{\mu}\,. \label{L_vs_S_6}%
\end{equation}
A factorization of a same type of vielbeins from $\bar{\theta}^{AB}$ would allows us to
write $d\mathcal{T}^{\left(  2p-1\right)  }\left(  \omega,\bar{\omega}\right)
$ in terms of the tensorial quantities (\ref{S_objects}) only, by means of the
identities (\ref{L_vs_S_4_0}-\ref{L_vs_S_5}). However, using (\ref{L_vs_S_5b}) in
(\ref{L_vs_S_6}) we get%
\begin{equation}
\bar{\theta}^{AB}=\left[  e_{\alpha}^{A}e^{B\gamma}\Gamma_{\mu\gamma}^{\alpha}-\eta^{BE}\left(  K^{-1}\right)_{\ C}^{A}K_{\ E}^{D}e_{\alpha}^{C}e_{D}^{\gamma}\bar{\Gamma}_{\mu\gamma
}^{\alpha} \right]  dx^{\mu}\,,
\label{L_vs_S_7}%
\end{equation}
which shows that, even when expressing $\bar{e}$ in terms of $K$ and $e$, we
cannot factorize the vielbeins $e$. Thus, one must discard the use of the
object $\bar{\theta}^{AB}$ in the Lorentz covariant constructions that can be
written in tensorial way\footnote{As another example, consider the Lorentz
invariant quantitiy $E^{A}E^{B}\eta_{AB}$, where $E^{A}$ is given in
(\ref{L_vs_S_3}) and where a tensorial product is assumed between $E^{A}$ and
$E^{B}$. Using (\ref{L_vs_S_5b}) we obtain
\par%
\[
E^{A}E^{B}\eta_{AB}=\left(  a^{2}g_{\mu\nu}+ab\left(  e_{\mu}^{A}e_{\nu}%
^{D}+e_{\nu}^{A}e_{\mu}^{D}\right)  \left(  K^{-1}\right)  _{\ D}^{B}\eta
_{AB}+b^{2}e_{\mu}^{C}e_{\nu}^{D}\bar{g}_{CD}\left(  x\right)  \right)
dx^{\mu}dx^{\nu}\,,%
\]
where we have also used the relation $\bar{g}_{CD}\left(  x\right)  =\left(
K^{-1}\right)  _{\ C}^{A}\left(  K^{-1}\right)  _{\ D}^{B}\eta_{AB}$ found in
Eq. (\ref{g_NL_2}) of the Appendix \ref{App_Rel_vielb}. We see that there is
no direct way to write $E^{A}E^{B}\eta_{AB}$ in terms of the tensorial objects
(\ref{S_objects}), i.e., as an expression free of vielbeins. Thus, one should
also discard the object $E^{A}$ from the Lorentz covariant constructions that
can be fully translated to tensorial language.}.

From the previous analysis, it is therefore clear that working only with the manifolds
$\mathcal{M}$ and $\mathcal{\bar{M}}$ is not sufficient to express the CW
theorem in tensorial language. In the next section we will introduce a \textit{hybrid}
manifold which will do the job.

\section{A tensorial Chern-Weil theorem\label{S_A_tens_version}}

\subsection{The \textit{hybrid} manifold $\mathcal{\breve{M}}$\label{Hybrid_space}}

Let us define the one-form $\tilde{\omega}_{\ B}^{A}$ as%
\begin{equation}
\tilde{\omega}_{\ B}^{A}=\tilde{\omega}_{\mu B}^{A}dx^{\mu}\,\equiv\left(
e_{\alpha}^{A}e_{B}^{\gamma}\bar{\Gamma}_{\mu\gamma}^{\alpha}+e_{\alpha}%
^{A}\partial_{\mu}e_{B}^{\alpha}\right)  dx^{\mu}\,. \label{hyb1}%
\end{equation}
A direct calculation shows that the transformation law of this object under a
local transformation $y^{\prime A}=\Lambda_{\ B}^{A}y^{B}$ is the same as Eq. (\ref{transf_w}). Thus,
$\tilde{\omega}$ is a spin connection allowing to define consistently
the covariant derivative of any Lorentz tensor\footnote{For example, if a Lorentz
vector $V^{A}$ transforms as $V^{\prime A}=\Lambda_{\ B}^{A^{\prime}}V^{B}$, then the covariant derivative $\tilde{D}%
V^{A}=dV^{A}+\tilde{\omega}_{\ B}^{A}V^{B}$ transforms as a vector too, i.e.,
$\tilde{D}^{\prime}V^{\prime A}=\Lambda_{\ B}^{A^{\prime}}\tilde{D}V^{B}$.}.
This connection has been recently introduced in Ref. \cite{Deruelle:2017xel}
and it has been named \textit{hybrid connection} because it depends on objects
that belong to different spaces: the vielbein $e$ associated with the manifold
$\mathcal{M}$ and the Christoffel symbol $\bar{\Gamma}$ of the manifold
$\mathcal{\bar{M}}$.

As a consequence, the difference between $\omega^{AB}=\eta^{BC}\omega_{\ C}^{A}$
and\ $\tilde{\omega}^{AB}=\eta^{BC}\tilde{\omega}_{\ C}^{A}$ is related with
the tensorial object $\Delta_{\mu\gamma}^{\alpha}=\Gamma_{\mu\gamma}^{\alpha
}-\bar{\Gamma}_{\mu\gamma}^{\alpha}$ defined in (\ref{L_vs_S_2}) as follows%
\begin{equation}
\tilde{\theta}^{AB}\equiv\omega^{AB}-\tilde{\omega}^{AB}=e_{\alpha}%
^{A}e_{\beta}^{B}\Delta_{\mu}^{\alpha\beta}dx^{\mu}\,,\text{ \ \ with
\ \ }\Delta_{\mu}^{\alpha\beta}=g^{\beta\gamma}\Delta_{\mu\gamma}^{\alpha}\,.
\label{hyb2}%
\end{equation}
The fact that two vielbeins of the same type can be factorized from
$\tilde{\theta}^{AB}$ (just as it happens, e.g., for $\Omega^{AB}=\frac{1}%
{2}e_{\alpha}^{A}e_{\beta}^{B}R_{\ \ \mu\nu}^{\alpha\beta}dx^{\mu}dx^{\nu}$)
is crucial to find a tensorial formulation of the Chern-Weil theorem. For
example, in the two dimensional case, if we replace the general connection
$\bar{\omega}$ by $\tilde{\omega}$ in the r.h.s of Eq. (\ref{cw2_gen_1}) and
use the identity (\ref{L_vs_S_4_0}) we get,%
\begin{equation}
d[\mathcal{T}^{\left(  1\right)  }(\tilde{\theta})]=d(\varepsilon_{AB}%
\tilde{\theta}^{AB})=d\left(  \varepsilon_{AB}e_{\alpha}^{A}e_{\beta}%
^{B}\Delta_{\nu}^{\alpha\beta}dx^{\nu}\right)  =\partial_{\mu}k_{\left(
1\right)  }^{\mu}d^{2}x\,, \label{hyb_ex}%
\end{equation}
where
\begin{equation}
k_{\left(  1\right)  }^{\mu}=\sqrt{-g}\delta_{\alpha\beta}^{\mu\nu}\Delta
_{\nu}^{\alpha\beta} \label{ten_CW_3b}%
\end{equation}
is a vector density that depends on the tensorial quantities (\ref{S_objects})
only: It is the Katz vector density \cite{Katz:1985} in $2$ dimensions.\\ 

Before showing how the l.h.s. of Eq. (\ref{cw2_gen_1}) would be modified if we change
$\bar{\omega}$ by $\tilde{\omega}$, a study of the geometric properties of the
hybrid connection is needed.\\

As shown in the Appendix \ref{Geom_prop_hyb}, $\tilde{\omega}^{AB}$ is not antisymmetric, hence the associated manifold $\mathcal{\tilde{M}}$ is not
metric compatible. On the other hand, the antisymmetrized object $\breve{\omega}_{\mu}%
^{AB}=\tilde{\omega}_{\mu}^{\left[  AB\right]  }$ has all the required
properties to define a Riemannian manifold $\mathcal{\breve{M}}$ (see Appendix
\ref{Geom_prop_hyb} for demonstrations): it transforms as a Lorentz spin connection
and is such that two vielbeins of the same type can be factorized from the
difference between $\omega^{AB}$ and $\breve{\omega}^{AB}$, that is
\begin{equation}
\breve{\theta}^{AB}\equiv\omega^{AB}-\breve{\omega}^{AB}=e_{\alpha}%
^{A}e_{\beta}^{B}\Delta_{\mu}^{\left[  \alpha\beta\right]  }dx^{\mu}\,,\text{
\ \ with \ \ }\Delta_{\mu}^{\alpha\beta}=g^{\beta\gamma}\Delta_{\mu\gamma
}^{\alpha}\,. \label{ant_hyb0}%
\end{equation}
Therefore, we introduce the \textit{antisymmetric hybrid }spin connection%
\begin{equation}
\breve{\omega}_{\mu B}^{A}\equiv\eta_{BC}\tilde{\omega}_{\mu}^{\left[
AC\right]  }=\frac{1}{2}\eta_{BC}\left(  e_{\alpha}^{A}e^{C\gamma}\bar{\Gamma}_{\mu\gamma}^{\alpha}-e_{\alpha}^{C}e^{A\gamma}\bar{\Gamma}_{\mu\gamma}^{\alpha}
+e_{\alpha}^{A}\partial_{\mu}e^{C\alpha}
-e_{\alpha}^{C}\partial_{\mu}e^{A\alpha}\right)  \,, \label{ant_hyb_def}%
\end{equation}
which can be associated with an auxiliary manifold $\mathcal{\breve{M}}$ with
metric $\breve{g}_{\mu\nu}$, affine connection $\breve{\Gamma}_{\mu\gamma
}^{\alpha}$ and vielbein $\breve{e}^{A}$ satisfying the basic relations
\begin{align}
\breve{g}_{\mu\nu}  &  =\breve{e}_{\mu}^{A}\breve{e}_{\nu}^{B}\eta
_{AB}\,,\label{ant_hyb1}\\
\breve{\omega}_{\mu B}^{A}  &  =\breve{e}_{\alpha}^{A}\breve{e}_{B}^{\gamma
}\breve{\Gamma}_{\mu\gamma}^{\alpha}+\breve{e}_{\alpha}^{A}\partial_{\mu
}\breve{e}_{B}^{\alpha}\,. \label{ant_hyb2}%
\end{align}
Thus, Eq. (\ref{ant_hyb2}) ensures that the Lorentz curvature and the torsion
two-forms
\[
\breve{\Omega}_{\ B}^{A}=d\breve{\omega}_{\ B}^{A}+\breve{\omega}_{\ C}%
^{A}\breve{\omega}_{\ B}^{C}=\frac{1}{2}\breve{\Omega}_{\ B\mu\nu}^{A}dx^{\mu
}dx^{\nu}\,,\ \ \ \ \ \ \breve{T}^{A}\equiv\breve{D}\breve{e}^{A}=\frac{1}%
{2}\breve{T}_{\mu\nu}^{A}dx^{\mu}dx^{\nu}\,,%
\]
are related with the Riemann and torsion tensors $\breve{R}_{\ \beta\mu\nu
}^{\alpha}=\partial_{\mu}\breve{\Gamma}_{\beta\nu}^{\alpha}-\ldots$ and
$\breve{T}_{\mu\nu}^{\lambda}=\breve{\Gamma}_{\mu\nu}^{\lambda}-\breve{\Gamma
}_{\nu\mu}^{\lambda}$ by%
\[
\breve{\Omega}_{\ B\mu\nu}^{A}=\breve{e}_{\alpha}^{A}\breve{e}_{B}^{\beta
}\breve{R}_{\ \beta\mu\nu}^{\alpha}\text{ \ \ and}\ \ \ \breve{T}_{\mu\nu}%
^{A}=\breve{e}_{\lambda}^{A}\breve{T}_{\mu\nu}^{\lambda}\,.
\]
We notice also that the Bianchi identities $\breve{D}\breve{\Omega}_{\ B}%
^{A}=0$ and $\breve{D}\breve{T}^{A}=\breve{\Omega}_{\ B}^{A}\breve{e}^{B}$ are
satisfied. Now, writing Eq. (\ref{ant_hyb_def}) as
\[
\breve{\omega}_{\mu}^{AB}=\breve{e}_{\alpha}^{A}\breve{\nabla}_{\mu}\breve
{e}^{B\alpha}=-\breve{e}^{B\alpha}\breve{\nabla}_{\mu}\breve{e}_{\alpha}%
^{A}\,,
\]
and using that by construction $\breve{\omega}_{\mu}^{AB}=-\breve{\omega}%
^{BA}$ we obtain $\breve{e}^{B\alpha}\breve{\nabla}_{\mu}\breve{e}_{\alpha
}^{A}=\breve{e}_{\alpha}^{B}\breve{\nabla}_{\mu}\breve{e}^{A\alpha}$ which
holds if and only if $\breve{\nabla}_{\mu}\breve{g}_{\alpha\beta}=0$.
Consequently, the manifold $\mathcal{\breve{M}}$ is metric compatible and
imposing torsionless condition we can ensure that $\breve{\Gamma}_{\mu\gamma
}^{\alpha}$ in (\ref{ant_hyb2}) is the Christoffel symbol, namely
$\breve{\Gamma}_{\mu\gamma}^{\alpha}=\breve{\Gamma}_{\mu\gamma}^{\alpha
}\left(  \breve{g},\partial\breve{g}\right)  $.

It is worth to point out that, usually, one considers the torsionless condition
$d\breve{e}^{A}+\breve{\omega}_{\ B}^{A}\breve{e}^{B}=0$ as a set of
$D^{2}\left(  D-1\right)  /2$ algebraic equations to solve the $D^{2}\left(
D-1\right)  /2$ components of the spin connection $\breve{\omega}_{\mu B}^{A}$
in terms of the vielbein, namely $\breve{\omega}_{\mu B}^{A}=\breve{\omega
}_{\mu B}^{A}\left(  \breve{e},\partial\breve{e}\right)  $. Here we
have the inverse problem. Given a set of functions $\breve{\omega}_{\mu B}%
^{A}$ that transform as a spin connection and that are calculated from the
definition (\ref{ant_hyb_def}), we use $\breve{\omega}_{\mu B}^{A}%
=\breve{\omega}_{\mu B}^{A}\left(  \breve{e},\partial\breve{e}\right)  $ as a
set of partial differential equations to determine the vielbein $\breve
{e}_{\mu}^{A}$, at least up to a Lorentz rotation. Having solved for the
vielbein, we can determine the metric $\breve{g}_{\mu\nu}=\breve{e}_{\mu}%
^{A}\breve{e}_{\nu}^{B}\eta_{AB}$\ and the Christoffel symbol $\breve{\Gamma
}_{\mu\gamma}^{\alpha}=\breve{\Gamma}_{\mu\gamma}^{\alpha}\left(  \breve
{g},\partial\breve{g}\right)  $. Then, one can check that Eq. (\ref{ant_hyb2})
is a consistency relation which must be satisfied\footnote{A similar situation
occurs in tensorial language. Given the metric tensor $\breve{g}_{\mu\nu}$ the
metricity condition $\nabla_{\lambda}\breve{g}_{\mu\nu}=0$ represents a set of
$D^{2}\left(  D+1\right)  /2$ algebraic equations to solve the $D^{2}\left(
D+1\right)  /2$ components of a torsionless connection $\breve{\Gamma}_{\mu
\nu}^{\lambda}$, which is symmetric in $\mu\nu$. The inverse process would be:
Given a set of functions $\breve{\Gamma}_{\mu\nu}^{\lambda}$ that transform as
an affine connection, the metricity condition can be regarded as a set of partial
differential equations to determine $\breve{g}_{\mu\nu}$. Integrability is
ensured by the fact that the symmetric connection $\breve{\Gamma}_{\mu\nu
}^{\lambda}$ allows to calculate the Riemann tensor $\breve{R}_{\ \beta\mu\nu
}^{\alpha}=\partial_{\mu}\breve{\Gamma}_{\beta\nu}^{\alpha}-\ldots$ which
characterizes univocally the geometry of a pseudo-Riemannain manifold
$\mathcal{\breve{M}}$. Therefore, the metric $\breve{g}_{\mu\nu}$ for which
$\breve{\Gamma}_{\mu\nu}^{\lambda}\left(  \breve{g},\partial\breve{g}\right)
$ is the Christoffel symbol can always be determined (up to a coordinate
transformation).}. 

As an example, let us consider the case where $\mathcal{M}_{D}$ and $\mathcal{\bar{M}}_{D}$ are static spherically symmetric
spacetimes, with metrics given by%
\begin{equation}
ds^{2}=-f^{2}\left(  r\right)  dt^{2}+\frac{1}{h^{2}\left(  r\right)  }%
dr^{2}+r^{2}\gamma_{nm}dx^{n}dx^{m}\,,\ \ \ d\bar{s}^{2}=-\bar{f}^{2}\left(
r\right)  dt^{2}+\frac{1}{\bar{h}^{2}\left(  r\right)  }dr^{2}+r^{2}%
\gamma_{nm}dx^{n}dx^{m}\,, \label{SSS_g_and_bar_g}%
\end{equation}
with $x^{n}=\phi_{1},\ldots\phi_{D-2}$ and $\gamma_{nm}$ is the metric of a
$(D-2)$-dimensional maximally symmetric space. It is direct to show that the
manifold $\mathcal{\breve{M}}_{D}$ has a metric%
\[
d\breve{s}^{2}=-\breve{f}^{2}\left(  r\right)  dt^{2}+\frac{1}{\breve{h}%
^{2}\left(  r\right)  }dr^{2}+r^{2} \gamma_{nm}dx^{n}dx^{m}\,,
\]
where
\[
\breve{h}=-\frac{1}{2\left(  D-2\right)  }\left(  \frac{1}{rh}\gamma^{nm}%
\bar{\Gamma}_{nm}^{r}-rh\Gamma_{nr}^{n}\right)  \,,\ \ \ \breve{f}\,=\int
\frac{\bar{f}\,^{\prime}}{2\breve{h}}\left(  \frac{f\,h}{\bar{f}}+\frac
{\bar{f}\,\bar{h}^{2}}{f\,h}\right)  \,dr\,,
\]
Thus, one can calculate the Christoffel symbol $\breve{\Gamma}_{\mu\gamma
}^{\alpha}\left(  \breve{g},\partial\breve{g}\right)  $, spin connection
$\breve{\omega}_{\mu B}^{A}$ and vielbein $\breve{e}_{\mu}^{A}$ associated
with this metric to show that, consistently, the tetrad postulate
(\ref{ant_hyb2}) is satisfied. It is also worth to mention that the
integration constant that appear after solving the differential equation for
$\ \breve{f}\,$\ can be easily fixed by demanding that $\mathcal{\breve{M}%
}\rightarrow\mathcal{\bar{M}}_{D}$ when $\mathcal{M}_{D}\rightarrow
\mathcal{\bar{M}}_{D}$. In the more specific case where $D=4$ and
$\mathcal{M}_{D}\,$, $\mathcal{\bar{M}}_{D}$ are respectively the Schwarzchild
and Minkowski metrics one gets
\[
d\breve{s}^{2}=-dt^{2}+\frac{\left(  1-\frac{2M}{r}\right)  }{\left(
1-\frac{M}{r}\right)  }dr^{2}+r^2(d\theta^2+\sin^2d\phi^2)\,.
\]

Having shown that $\mathcal{\breve{M}}$ is not a new independent manifold,
because its geometry can be completely characterized with the geometric
quantities of $\mathcal{M}$ and $\mathcal{\bar{M}}$, we are now able to give a tensorial version of the CW theorem.

\subsection{Chern-Weil theorem for the hybrid manifold}

Given a pair of pseudo-Riemannian manifolds $(\mathcal{M},\mathcal{\bar{M}})$
the hybrid connection $\breve{\omega}$ defined in Eq. (\ref{ant_hyb_def})
allows to construct a third auxiliary pseudo-Riemannian manifold
$\mathcal{\breve{M}}$ whose geometry is completely determined in terms of the
first two, namely $\mathcal{\breve{M}}=\mathcal{\breve{M}}_{(\mathcal{M}%
,\mathcal{\bar{M}})}\,$. Considering first the two-dimensional case, the Chern-Weil theorem for the manifolds $\mathcal{M}$ and $\mathcal{\breve{M}}$
reads%
\begin{equation}
\mathcal{E}_{2}\left(  \Omega\right)  -\mathcal{E}_{2}(\breve{\Omega
})=d[\mathcal{T}^{\left(  1\right)  }(\breve{\theta})]\,, \label{ten_CW_1}%
\end{equation}
where $\mathcal{E}_{2}\left(  \Omega\right)  =\varepsilon_{AB}\Omega^{AB}\,$,
$\mathcal{E}_{2}(\breve{\Omega})=\varepsilon_{AB}\breve{\Omega}^{AB}\,$,
$\mathcal{T}^{\left(  1\right)  }(\breve{\theta})=\varepsilon_{AB}%
\breve{\theta}^{AB}\,$, $\breve{\theta}^{AB}=\omega^{AB}-\breve{\omega}%
^{AB}\,$. The l.h.s. of (\ref{ten_CW_1}) can be easily written in terms of
tensorial objects of $\mathcal{M}$ and $\mathcal{\breve{M}}$ while the
translation of the r.h.s. using the tetrad postulates for $\omega\,$\ and
$\breve{\omega}$ only is problematic (see Eq. (\ref{cw2_gen_2}) of the Section
\ref{S_Issues_A}). On the other hand, if we use the definition of the hybrid connection $\breve{\omega}$ given by Eq. (\ref{ant_hyb_def}), the
r.h.s. of (\ref{ten_CW_1}) can be written as a tensorial expression with
respect to the pair $(\mathcal{M},\mathcal{\bar{M}})$ instead of
$(\mathcal{M},\mathcal{\breve{M}})$. Indeed, we have%
\begin{equation}
d[\mathcal{T}^{\left(  1\right)  }(\breve{\theta})]=d(\varepsilon_{AB}%
\breve{\theta}^{AB})=d(\varepsilon_{AB}e_{\alpha}^{A}e_{\beta}^{B}\Delta_{\nu
}^{\alpha\beta}dx^{\nu})=\partial_{\mu}k_{\left(  1\right)  }^{\mu}d^{2}x\,,
\label{ten_CW_3}%
\end{equation}
where $k_{\left(  1\right)  }^{\mu}=\sqrt{-g}\delta_{\alpha\beta}^{\mu\nu
}\Delta_{\nu}^{\alpha\beta}$ is the Katz vector density in $2$ dimensions.
Thus, Eq. (\ref{ten_CW_1}) reads%
\[
\left(  \sqrt{-g}R-\sqrt{-\breve{g}}\,\breve{R}\right)  d^{2}x=\partial_{\mu
}k_{\left(  1\right)  }^{\mu}d^{2}x\,.
\]
Denoting by $\mathcal{E}_{2}\left[  \mathcal{M}\right]  =\sqrt{-g}Rd^{2}x$ and
$\mathcal{E}_{2}[\mathcal{\breve{M}}]=\sqrt{-\breve{g}}\breve{R}\,d^{2}x$ the
tensorial expression for the topological terms associated with
$\mathcal{M}$ and $\mathcal{\breve{M}}$, and denoting by $k_{\left(  1\right)
}^{\mu}\left[  \mathcal{M},\mathcal{\bar{M}}\right]  $ the vector density
(\ref{ten_CW_3b}) which depends on the tensorial objects of $\mathcal{M}$ and
$\mathcal{\bar{M}}$, the tensorial version that we have obtained for the
Chern-Weil theorem has the following schematical structure,%
\begin{equation}
\mathcal{E}_{2}\left[  \mathcal{M}\right]  -\mathcal{E}_{2}[\mathcal{\breve
{M}}_{\left(  \mathcal{M},\mathcal{\bar{M}}\right)  }]=\partial_{\mu
}(k_{\left(  1\right)  }^{\mu}\left[  \mathcal{M},\mathcal{\bar{M}}\right]
)d^{2}x\,. \label{ten_CW_4}%
\end{equation}

This result can be extended for any pair of given $2p$-dimensional Riemannian
manifolds $(\mathcal{M},\mathcal{\bar{M}})$. After constructing the auxiliary
manifold $\mathcal{\breve{M}}$ with the hybrid connection (\ref{ant_hyb_def})
the Chern-Weil theorem for the pair $(\mathcal{M},\mathcal{\breve{M}%
}_{(\mathcal{M},\mathcal{\bar{M}})})$ is given by%
\begin{equation}
\mathcal{E}_{2p}\left(  \Omega\right)  -\mathcal{E}_{2p}(\breve{\Omega
})=d[\mathcal{T}^{\left(  2p-1\right)  }(\breve{\theta},\Omega,\breve{\Omega
})]\,,\text{ \ \ with \ \ }\breve{\theta}^{AB}\equiv\omega^{AB}-\breve{\omega
}^{AB}\,, \label{CW_7}%
\end{equation}
where%
\begin{align}
\mathcal{E}_{2p}\left(  \Omega\right)   &  \equiv\varepsilon_{A_{1}\ldots
A_{2p}}\Omega^{A_{1}A_{2}}\cdots\Omega^{A_{2p-1}A_{2p}}=\frac{1}{2^{p}}%
\sqrt{-g}\delta_{\alpha_{1}\ldots\alpha_{2p}}^{\mu_{1}\ldots\mu_{2p}}%
R_{\mu_{1}\mu_{2}}^{\alpha_{1}\alpha_{2}}\cdots R_{\mu_{2p-1}\mu_{2p}}%
^{\alpha_{2p-1}\alpha_{2p}}d^{2p}x\,,\nonumber\\
\mathcal{E}_{2p}(\breve{\Omega})  &  \equiv\varepsilon_{A_{1}\ldots A_{2p}%
}\breve{\Omega}^{A_{1}A_{2}}\cdots\breve{\Omega}^{A_{2p-1}A_{2p}}=\frac
{1}{2^{p}}\sqrt{-\breve{g}}\delta_{\alpha_{1}\ldots\alpha_{2p}}^{\mu_{1}%
\ldots\mu_{2p}}\breve{R}_{\mu_{1}\mu_{2}}^{\alpha_{1}\alpha_{2}}\cdots
\breve{R}_{\mu_{2p-1}\mu_{2p}}^{\alpha_{2p-1}\alpha_{2p}}d^{2p}x\,, \label{CW_6}%
\end{align}
are the corresponding Euler terms and
\begin{equation}
\mathcal{T}^{\left(  2p-1\right)  }(\breve{\theta},\Omega,\breve{\Omega
})=p\int_{0}^{1}dt\varepsilon_{A_{1}\ldots A_{2p}}\breve{\theta}^{A_{1}A_{2}%
}\Omega_{\left(  t\right)  }^{A_{3}A_{4}}\cdots\,\Omega_{\left(  t\right)
}^{A_{2p-1}A_{2p}}\, \label{CW_8}%
\end{equation}
is the transgression form with $\Omega_{\left(  t\right)  }^{AB}%
=d\omega_{\left(  t\right)  }^{AB}+\omega_{\left(  t\right)  C}^{A}%
\omega_{\left(  t\right)  }^{CB}$ and $\omega_{\left(  t\right)  }^{AB}%
=\breve{\omega}^{AB}+t\breve{\theta}^{AB}$ is a connection which interpolates
between $\breve{\omega}^{AB}$ and $\omega^{AB}$.

From the different alternative expressions that the interpolating curvature
may have (see (\ref{int_curv_a}-\ref{int_curv_c})),%
\begin{equation}
\Omega_{\left(  t\right)  }^{AB}=\Omega^{AB}+\left(  t-1\right)
D\breve{\theta}^{AB}+\left(  t-1\right)  ^{2}\breve{\theta}_{\ C}^{A}%
\breve{\theta}^{CB} \label{hyb_interp_Curv_c}%
\end{equation}
is useful to translate the exterior derivative of the
transgression (\ref{CW_8}) to tensorial language. Indeed, using Eq.
(\ref{ant_hyb0}), the relation $g_{\gamma\lambda}=\eta_{CD}e_{\gamma}%
^{C}e_{\lambda}^{D}$ and
\begin{equation}
D\breve{\theta}^{AB}=D_{\mu}\breve{\theta}_{\nu}^{AB}dx^{\mu}dx^{\nu
}=e_{\alpha}^{A}e_{\beta}^{B}\nabla_{\mu}\Delta_{\nu}^{\left[  \alpha
\beta\right]  }dx^{\mu}dx^{\nu},\ \ \ \text{with}\ \ \ \Delta_{\mu}%
^{\alpha\beta}=g^{\beta\gamma}\Delta_{\mu\gamma}^{\alpha}\,, \label{cww3}%
\end{equation}
which can be proved using the tetrad postulate plus the torsionless condition,
we get%
\begin{equation}
\Omega_{\left(  t\right)  }^{AB}=e_{\alpha}^{A}e_{\beta}^{B}\left(  \frac
{1}{2}R_{\ \ \mu\nu}^{\alpha\beta}+\left(  t-1\right)  \nabla_{\mu}\Delta
_{\nu}^{\left[  \alpha\beta\right]  }+\left(  t-1\right)  ^{2}g_{\gamma
\lambda}\Delta_{\mu}^{\left[  \alpha\gamma\right]  }\Delta_{\nu}^{\left[
\lambda\beta\right]  }\right)  dx^{\mu}dx^{\nu}\,. \label{cww2}%
\end{equation}
Thus, vielbeins of the same type can be factorized from $\breve{\theta}%
^{A_{1}A_{2}}$ and each interpolating curvature in (\ref{CW_8}) so the use of
the identity (\ref{df_3}) allows to write%
\begin{equation}
d[\mathcal{T}^{\left(  2p-1\right)  }(\breve{\theta},\Omega,\breve{\Omega
})]=\partial_{\mu}k_{\left(  p\right)  }^{\mu}d^{2p}x\,, \label{cww4}%
\end{equation}
where%
\begin{align}
k_{\left(  p\right)  }^{\mu}  &  =\sqrt{-g}\,p\int_{0}^{1}dt\,\delta
_{\alpha_{1}\ldots\alpha_{2p}}^{\mu\nu_{2}\ldots\nu_{2p}}\Delta_{\nu_{2}%
}^{\alpha_{1}\alpha_{2}}\left(  \frac{1}{2}R_{\ \ \nu_{3}\nu_{4}}^{\alpha
_{3}\alpha_{4}}+\left(  t-1\right)  \nabla_{\nu_{3}}\Delta_{\nu_{4}}%
^{\alpha_{3}\alpha_{4}}+\left(  t-1\right)  ^{2}g_{\gamma_{1}\lambda_{1}%
}\Delta_{\nu_{3}}^{\left[  \alpha_{3}\gamma_{1}\right]  }\Delta_{\nu_{4}%
}^{\left[  \lambda_{1}\alpha_{4}\right]  }\right)  \times\cdots\nonumber\\
&  \,\cdots\times\left(  \frac{1}{2}R_{\ \ \nu_{2p-1}\nu_{2p}}^{\alpha
_{2p-1}\alpha_{2p}}+\left(  t-1\right)  \nabla_{\nu_{2p-1}}\Delta_{\nu_{2p}%
}^{\alpha_{2p-1}\alpha_{2p}}+\left(  t-1\right)  ^{2}g_{\gamma_{p}\lambda_{p}%
}\Delta_{\nu_{2p-1}}^{\left[  \alpha_{2p-1}\gamma_{p}\right]  }\Delta
_{\nu_{2p}}^{\left[  \lambda_{p}\alpha_{2p-1}\right]  }\right)  \label{CW_9}%
\end{align}
is a vector density that is completely characterized by the manifolds
$\mathcal{M}$ and $\mathcal{\bar{M}}$, i.e., $k_{\left(  p\right)  }^{\mu
}=k_{\left(  p\right)  }^{\mu}\left[  \mathcal{M},\mathcal{\bar{M}}\right]  $.
Therefore, the tensorial version of the CW theorem that we have constructed
can be written as,
\begin{equation}
\frac{1}{2^{p}}\sqrt{-g}\,\delta_{\alpha_{1}\ldots\alpha_{2p}}^{\mu_{1}%
\ldots\mu_{2p}}\left(  R_{\mu_{1}\mu_{2}}^{\alpha_{1}\alpha_{2}}\cdots
R_{\mu_{2p-1}\mu_{2p}}^{\alpha_{2p-1}\alpha_{2p}}-\breve{R}_{\mu_{1}\mu_{2}%
}^{\alpha_{1}\alpha_{2}}\cdots\breve{R}_{\mu_{2p-1}\mu_{2p}}^{\alpha
_{2p-1}\alpha_{2p}}\right)  d^{2p}x=\partial_{\mu}k_{\left(  p\right)  }^{\mu
}d^{2p}x\,, \label{CW_tens}%
\end{equation}
which has the following structure%
\begin{equation}
\mathcal{E}_{2p}\left[  \mathcal{M}\right]  -\mathcal{E}_{2p}[\mathcal{\breve
{M}}_{\left(  \mathcal{M},\mathcal{\bar{M}}\right)  }]=\partial_{\mu
}(k_{\left(  p\right)  }^{\mu}\left[  \mathcal{M},\mathcal{\bar{M}}\right]
)d^{2p}x\,, \label{CW_Scheme}%
\end{equation}
where $\mathcal{E}_{2p}\left[  \mathcal{M}\right]  $ and $\mathcal{E}%
_{2p}[\mathcal{\breve{M}}]$ denotes the tensorial expression of the Euler terms (\ref{CW_6}).\\

Eq. (\ref{CW_9}-\ref{CW_Scheme}) is the result we aimed at : to write the Chern-Weil theorem is terms of purely spacetime tensorial quantities.
As we have seen in detail, this requires the explicit introduction of a background manifold, $\cal{\bar M}$ whose role is (1) to construct the
spacetime tensors $\Delta=\Gamma-\bar\Gamma$ which are essential in the definition of the vector $k^{\mu}_{(p)}$, see (4.20); (2) to construct the manifold $\breve M$
such that the divergence of the vector $k^{\mu}_{(p)}$ is the difference of the topological terms of ${\cal M}$ and $\cal{\breve M}$.

In the case we choose $\cal {\bar M}$ to be a product manifold (whose metric can be written as $d\bar{s}^2=dr^2+\bar h_{ij}(x^i)dx^i dx^j$ so that the extrinsic curvatures of the $r=const$ hypersurfaces are zero) its topological term vanishes  (see \cite{Eguchi:1980jx}), as well as that of $\cal{\breve M}$ (for the same reasons, as can be easily shown from of the definition of the hybrid connection (\ref{ant_hyb_def})).
This shows explicitely that the Einstein-Hilbert, Gauss-Bonnet and in general the Lovelock terms reduce, in the critical dimension $D=2p$, to the divergence of a vector density constructed with spacetime tensors, ${\cal E}_{2p}[{\cal M}]=\partial_\mu k^\mu_{(p)}$.
If the product manifold $\cal {\bar M}$ is also cobordant, which means that a specific surface $r=r_0$ coincides with the boundary of $\cal M$, then the component $k^r_{(p)}$ normal to the boundary coincides with Myers' boundary term (see \cite{Deruelle:2017xel}).
However, this does not mean that there is no underlying structure, since the vector $k^\mu_{(p)}$ still depends explicitely on $\cal {\bar M}$.
In the generic case, the topological term of $\cal{\breve M}$ does not vanish and represents a topological obstruction to write the Euler term of $\cal M$ as a divergence of the vector density $k^{\mu}_{(p)}$.

In all cases (whether $\cal {\bar M}$ is an arbitrary background or product manifold) the tensorial translation of the Chern-Weil theorem requires the introduction
of an the extra structure $\cal {\bar M}$. In fact this should not come as a surprise : for example, Horndeski, in his proof \cite{Horndeski} that 
${\cal E}_{2p}[{\cal M}] \propto \partial_\mu V^\mu_{(p)}$, where
\begin{equation}
V^{\mu}_{(p)} = V^{\mu}_{(p)} (g_{\mu \nu}, R^{\mu}_{\nu \alpha \beta}, U^{\mu},  \nabla_{\nu}U^{\mu})\,,
\end{equation}
depends also on an extra structure, namely $U^{\mu}$ which is an arbitrary non-null contravariant vector (see the explicit expression of $V^\mu_{(p)}$ in Appendix \ref{Myers_from_Horndeski}).

\section{Final remarks\label{Final}}

Given two-pseudo Riemannian manifolds $\mathcal{M}$ and $\mathcal{\bar{M}}$ we
have introduced an auxiliary manifold $\mathcal{\breve{M}}$ whose geometry is
completely determined by the first two and that allows to construct the
tensorial version of the Chern-Weil theorem (\ref{CW_tens}). This expression
states that the difference of the Euler terms of
$\mathcal{M}$ and $\mathcal{\breve{M}}_{\left(  \mathcal{M},\mathcal{\bar{M}%
}\right)  }$ is the divergence of the vector density (\ref{CW_9}) which
is constructed with objects that are tensorial with respect to $\mathcal{M}$
and $\mathcal{\bar{M}}$.

As we will see in \cite{KatzLovelock} (see also \cite{Deruelle:2017xel} for
the Gauss-Bonnet case), the tensorial version of the CW theorem presented in this work is the one that must be used (together with a dimensional
continuation procedure) to generalize the procedure developed by
Katz, Bicak and Lynden-Bell (KBL) \cite{Katz:1985},\cite{Katz:1996nr}\ to
calculate conserved charges in a generic Lovelock theory. There the manifolds
$\mathcal{M}$ and $\mathcal{\bar{M}}$ are interpreted as the dynamical and
background manifolds, while the hybrid manifold $\mathcal{\breve{M}}$ is just
an auxiliary manifold allowing us to write the KBL vector in the vielbein
formalism and to give a proof for the Dirichlet problem in Lovelock gravity.

\section{Acknowledgment}

NM was supported by a Becas-Chile postdoctoral grant of CONICYT. The work of
RO is funded in part by FONDECYT Grant No. 1170765, UNAB Grant DI-1336-16/R
and CONICYT Grant DPI 20140115. The authors are grateful to Gregory Horndeski for very useful correspondence, as well as to Milutin Blagojevic, Branislav Cvetkovic, F\'{e}lix Juli\'{e} and Olivera Miskovic for
useful comments.

\appendix

\section{Myers' terms from Horndeski divergences\label{Myers_from_Horndeski}}

Myers' boundary terms \cite{Myers:1987yn} in the critical dimension $D=2p$ are given by%
\[
I_{\text{Myers}}=\int_{\partial\mathcal{M}_{D}}d^{2p-1}x \, \beta^{\left(p\right)  }\,,
\]
with%
\begin{align*}
\beta^{\left(  p\right)  } &  =-2\sqrt{-h}\int_{0}^{1}dt \, \delta_{i_{1}\ldots
i_{2p-1}}^{j_{1}\ldots j_{2p-1}}K_{j_{1}}^{i_{1}}\left(  \frac{1}%
{2}\mathcal{R}_{j_{2}j_{3}}^{i_{2}i_{3}}-t^{2}K_{j_{2}}^{i_{2}}K_{j_{3}%
}^{i_{3}}\right)  \times\cdots\times\left(  \frac{1}{2}\mathcal{R}%
_{j_{2p-2}j_{2p-1}}^{i_{2p-2}i_{2p-1}}-t^{2}K_{j_{2p-2}}^{i_{2p-2}}%
K_{j_{2p-1}}^{i_{2p-1}}\right)  \\
&  =-2p\sqrt{-h}\left[  \frac{1}{2^{p+1}}%
{\displaystyle\sum_{s=0}^{p-1}}
\frac{4^{p-s}\left(  p-1\right)  !}{s!\left(  2p-2s-1\right)  !!} \, \delta_{i_{1}\ldots i_{2p-1}}^{j_{1}\ldots j_{2p-1}}R_{j_{1}j_{2}}^{i_{1}i_{2}%
}\cdots R_{j_{2s-1}j_{2s}}^{i_{2s-1}i_{2s}}K_{j_{2s+1}}^{i_{2s+1}}\cdots
K_{j_{2p-1}}^{i_{2p-1}}\right]\,,
\end{align*}
where $\mathcal{R}_{kl}^{ij}=\mathcal{R}_{kl}^{ij}\left(  h\right)  $ is the
intrinsic curvature of the boundary, $R_{kl}^{ij}=R_{kl}^{ij}\left(  g\right)
$ are the boundary components of the bulk curvature and the coefficients in the
last equality comes after performing the integration in the parameter $t$
(further details can be found, e.g., in Ref. \cite{Miskovic:2007mg}). In particular, the double factorial
is defined as%
\[
n!!=\left\{
\begin{tabular}
[c]{ll}%
$%
{\displaystyle\prod_{k=1}^{n/2}}
\left(  2k\right)  $ & , for $n$ even\,,\\
$%
{\displaystyle\prod_{k=1}^{\left(  n+1\right)  /2}}
\left(  2k-1\right)  $ & , for $n$ odd \,.
\end{tabular}
\ \ \right.
\]

On the other hand, in Ref. \cite{Horndeski} Horndeski has shown explicitely
that in the critical dimensions the Lovelock densities are given by a
divergence, namely%
\begin{equation}
2^{p}\int_{\mathcal{M}_{D}}d^{2p}x \, \mathcal{E}_{2p}[\mathcal{M}]=\int_{\mathcal{M}_{D}}d^{2p}x\sqrt{-g} \, \delta_{\nu_{1}\ldots\nu_{2p}}^{\mu
_{1}\ldots\mu_{2p}}R_{\mu_{1}\mu_{2}}^{\nu_{1}\nu_{2}}\cdots R_{\mu_{2p-1}%
\mu_{2p}}^{\nu_{2p-1}\nu_{2p}}=\int_{\mathcal{M}_{D}}d^{2p}x \, \partial_{\mu
}V_{\left(  p\right)  }^{\mu}\,,
\label{Hornd}%
\end{equation}
with $V_{\left(  p\right)  }^{\mu}$ being the following vector density
\[
V_{\left(  p\right)  }^{\mu}=\sqrt{-g} \, {\displaystyle\sum_{k=0}^{p-1}}
\frac{C_{k}}{\rho^{k+1}} \, \delta_{\nu\nu_{1}\ldots\nu_{2p-1}}^{\mu\mu_{1}%
\ldots\mu_{2p-1}}U^{\nu}\nabla_{\mu_{1}}U^{\nu_{1}}\cdots\nabla_{\mu_{2k+1}%
}U^{\nu_{2k+1}}R_{\mu_{2k+2}\mu_{2k+3}}^{\nu_{2k+2}\nu_{2k+3}}\cdots
R_{\mu_{2p-2}\mu_{2p-1}}^{\nu_{2p-2}\nu_{2p-1}}\,,
\]
where $\rho=U_{\mu}U^{\mu}$, $C_{0}=-4p$ and%
\[
C_{k}=-4^{k+1}p%
{\displaystyle\prod_{q=0}^{k-1}}
\left(  \frac{p-q-1}{2q+3}\right)\,.
\]

Here we show that if we use a radial foliation, with Gauss normal
coordinates given by $ds^{2}= dr^{2}+h_{ij}\left(
r,x^{i}\right)  dx^{i}dx^{j}$ and if we chose the arbitrary vector $U^{\mu}$
to be the normal vector of the surfaces $r=const$, namely $U^{\mu}=\left(1,\overrightarrow{0}\right)  $, then the Eq. (\ref{Hornd}) reproduces the Myers
boundary term as%
\[
\int_{\mathcal{M}_{D}}d^{2p}x \, \partial_{\mu}V_{\left(  p\right)  }^{\mu}%
=2^{p}I_{\text{Myers}}\,.
\]
For doing so, we first write the Myers term as%
\begin{align}
2^{p}I_{\text{Myers}} &  =-p\int_{\partial\mathcal{M}_{D}}d^{2p-1}x\sqrt{-h} \, {\displaystyle\sum_{s=0}^{p-1}}
\frac{4^{p-s}\left(  p-1\right)  !}{s!\left(  2p-2s-1\right)  !!} \, \delta
_{i_{1}\ldots i_{2p-1}}^{j_{1}\ldots j_{2p-1}}R_{j_{1}j_{2}}^{i_{1}i_{2}%
}\cdots R_{j_{2s-1}j_{2s}}^{i_{2s-1}i_{2s}}K_{j_{2s+1}}^{i_{2s+1}}\cdots
K_{j_{2p-1}}^{i_{2p-1}}\nonumber\\
&  =\int_{\partial\mathcal{M}_{D}}d^{2p-1}x\sqrt{-h}\left[  -4p \, \delta
_{i_{1}\ldots i_{2p-1}}^{j_{1}\ldots j_{2p-1}}R_{j_{1}j_{2}}^{i_{1}i_{2}%
}\cdots R_{j_{2s-1}j_{2s}}^{i_{2s-1}i_{2s}}K_{j_{2s+1}}^{i_{2s+1}}\cdots
K_{j_{2p-1}}^{i_{2p-1}}\right.  \nonumber\\
&  \left.  -%
{\displaystyle\sum_{s=0}^{p-2}}
\frac{4^{p-s}p\left(  p-1\right)  !}{s!\left(  2p-2s-1\right)  !!} \, \delta_{i_{1}\ldots i_{2p-1}}^{j_{1}\ldots j_{2p-1}}R_{j_{1}j_{2}}^{i_{1}%
i_{2}}\cdots R_{j_{2s-1}j_{2s}}^{i_{2s-1}i_{2s}}K_{j_{2s+1}}^{i_{2s+1}}\cdots
K_{j_{2p-1}}^{i_{2p-1}}\right]\,.
\label{M1}%
\end{align}

Now, using the Gauss theorem in this radial foliation, we can write the Horndeski term as%
\begin{align*}
\int_{\mathcal{M}_{D}}d^{2p}x \, \partial_{\mu}V_{\left(  p\right)  }^{\mu}  &
=\int_{\partial\mathcal{M}_{D}}d^{2p-1}x \, U_{\mu}V_{\left(  p\right)  }^{\mu
}= \int_{\partial\mathcal{M}_{D}}d^{2p-1}x \, V_{\left(  p\right)  }^{r}\\
&  =\int_{\partial\mathcal{M}_{D}}d^{2p-1}x\sqrt{-h} \, {\displaystyle\sum_{k=0}^{p-1}} C_{k} \, \delta_{i_{1}\ldots i_{2p-1}}^{j_{1}\ldots j_{2p-1}}K_{j_{1}}^{i_{1}%
}\cdots K_{j_{2k+1}}^{i_{2k+1}}R_{j_{2k+2}j_{2k+3}}^{i_{2k+2}i_{2k+3}}\cdots
R_{j_{2p-2}j_{2p-1}}^{i_{2p-2}i_{2p-1}}\,,
\end{align*}
where $V_{\left(  p\right)  }^{r}$ has been calculated using $\rho=1$, $U^{r}=1$, $\delta_{ri_{1}\ldots i_{2p-1}%
}^{rj_{1}\ldots j_{2p-1}}=\delta_{i_{1}\ldots i_{2p-1}}^{j_{1}\ldots j_{2p-1}%
}$ and $\nabla_{j}U^{i}=K_{j}^{i}$. 
Rearranging the indices, the Horndeski term
can we rewritten as%
\begin{align}
\int_{\mathcal{M}_{D}}d^{2p}x \, \partial_{\mu}V_{\left(  p\right)  }^{\mu}  &
=\int_{\partial\mathcal{M}_{D}}d^{2p-1}x\sqrt{-h} \, {\displaystyle\sum_{s=0}^{p-1}} C_{p-s-1}\delta_{i_{1}\ldots i_{2p-1}}^{j_{1}\ldots j_{2p-1}}R_{j_{1}j_{2}%
}^{i_{1}i_{2}}\cdots R_{j_{2s-1}j_{2s}}^{i_{2s-1}i_{2s}}K_{j_{2s+1}}^{i_{2s+1}}\cdots K_{j_{2p-1}}^{i_{2p-1}}\nonumber\\
&  =\int_{\partial\mathcal{M}_{D}}d^{2p-1}x\sqrt{-h}\left[  -4p \, \delta
_{i_{1}\ldots i_{2p-1}}^{j_{1}\ldots j_{2p-1}}R_{j_{1}j_{2}}^{i_{1}i_{2}%
}\cdots R_{j_{2p-3}j_{2p-2}}^{i_{2p-3}i_{2p-2}}K_{j_{2p-1}}^{i_{2p-1}}\right.
\nonumber\\
&  \left.
{\displaystyle\sum_{s=0}^{p-2}}
C_{p-s-1}\delta_{i_{1}\ldots i_{2p-1}}^{j_{1}\ldots j_{2p-1}}R_{j_{1}j_{2}%
}^{i_{1}i_{2}}\cdots R_{j_{2s-1}j_{2s}}^{i_{2s-1}i_{2s}}K_{j_{2s+1}}%
^{i_{2s+1}}\cdots K_{j_{2p-1}}^{i_{2p-1}}\right] \,,
\label{H1}%
\end{align}
where one can directly check that the first terms in (\ref{M1}) and (\ref{H1}) coincide, while the rest of the terms coincides
because%
\[
C_{p-s-1}=-4^{p-s}p%
{\displaystyle\prod_{q=0}^{p-s-2}}
\left(  \frac{p-q-1}{2q+3}\right)  =-4^{p-s}p \, \frac{\left(  p-1\right)
!}{s!\left(  2p-2s-1\right)  !!}\,. \ \ QED.
\]

\section{Transgression forms\label{App_Tran}}

Let $\mathcal{G}=\left\{  T_{M}\right\}  $ be a Lie algebra and $A=A_{\mu}%
^{M}T_{M}dx^{\mu}$ a Lie valued one-form gauge connection. This means that
under a gauge transformation characterized by a group element $g=\exp\left(
g^{M}T_{M}\right)  $ (the parameters $g^{M}$ being coordinates in the Lie
group manifold $G$) the connection transforms as
\[
A\rightarrow gAg^{-1}+gdg^{-1}\,.%
\]
If $Q$ is a $r$-form transforming as $Q\rightarrow gQg^{-1}$ under a gauge
transformation, then the covariant derivative with respect to $A$ is defined
by $\mathrm{D}Q=dQ+\left[  A,Q\right]  $ (with $\left[  \ ,\ \right]  $ being
the commutator) and transforms as $\mathrm{D}Q\rightarrow g\left(
\mathrm{D}Q\right)  g^{-1}$.

The strength field two-form is defined as $F=dA+AA\,$, where for shortness we
omit the wedge product between differential forms. Thus, by construction the
strengh field transform as $F\rightarrow gFg^{-1}$ and satisfies the Bianchi
identity, $\mathrm{D}F=0$. It is also direct to show that the symmetrized
trace of a given strength field power, denoted by $\left\langle F^{p}\right\rangle $, is invariant under gauge transformations. This can be
directly shown using the properties of the symmetrized the trace and the wedge
product. Hence, $\left\langle F^{p}\right\rangle $ is usually called a
\textit{topological term}.

A transgression form is defined by the Chern-Weyl theorem, which states that
if $A$\ and $\bar{A}$\ are two gauge connections valued on the same algebra
with strength fields $F$ and $\bar{F}$, then $\left\langle F^{p}\right\rangle
$ and $\left\langle \bar{F}^{p}\right\rangle $ are closed forms, i.e.,
\begin{equation}
d\left\langle F^{p}\right\rangle =d\left\langle \bar{F}^{p}\right\rangle =0\,
\label{T1}
\end{equation}
and
\begin{equation}
\left\langle F^{p}\right\rangle -\left\langle \bar{F}^{p}\right\rangle
=dT^{\left(  2p-1\right)  }\left(  A,\bar{A}\right)  \,, \label{T2}%
\end{equation}
where%
\begin{equation}
T^{\left(  2p-1\right)  }\left(  A,\bar{A}\right)  =p\int_{0}^{1}%
dt\left\langle \bar{\theta}F_{t}^{p-1}\right\rangle \,, \label{T3}%
\end{equation}
with $\bar{\theta}=A-\bar{A}$, $F_{t}=dA_{t}+A_{t}A_{t}$ and where $A_{t}%
=\bar{A}+t\bar{\theta}$ is a connection interpolating between $\bar{A}$ and
$A$. Eq. (\ref{T1}) states that both topological terms are closed forms,
whilst Eq. (\ref{T2}) tells that their difference is an exact form defined in
(\ref{T3}) by the $\left(  2p-1\right)  $-form $T^{\left(  2p-1\right)  }$,
which is known as a \textit{transgression form}.

In this construction, we frequently use the property $d\left\langle
Q\right\rangle =\left\langle \mathrm{D}Q\right\rangle $ when $Q$ is a
covariant object transforming as $Q\rightarrow gQg^{-1}$. Using $\mathrm{\bar
{D}}\bar{\theta}=\mathrm{D}\bar{\theta}-2\bar{\theta}^{2}$ and $F_{1}=F$, the
interpolating curvature can be written in the following three alternative
forms,%
\begin{align}
F_{t}  &  =\bar{F}+t\mathrm{\bar{D}}\bar{\theta}+t^{2}\bar{\theta}%
^{2}\,,\label{Ft_1}\\
&  =F+\left(  t-1\right)  \mathrm{D}\bar{\theta}+\left(  t^{2}-2t+1\right)
\bar{\theta}^{2}\,,\label{Ft_2}\\
&  =tF+\left(  1-t\right)  \bar{F}-t\left(  1-t\right)  \bar{\theta}^{2}\,.
\label{Ft_3}%
\end{align}

Let us consider now the case where the symmetry is given by the Lorentz
algebra with generators $\left\{  T_{M}\right\}  =\left\{  J_{AB}\right\}  $
satisfying,%
\begin{equation}
\left[  J_{AB},J_{CD}\right]  =\eta_{BC}J_{AD}-\eta_{AC}J_{BD}+\eta_{AD}%
J_{BC}-\eta_{BD}J_{AC}\,. \label{L_com}%
\end{equation}
In this case we have,%
\begin{align}
A  &  =\frac{1}{2}\omega^{AB}J_{AB}\,,\ \ \bar{A}=\frac{1}{2}\bar{\omega}%
^{AB}J_{AB}\,,\ \ \bar{\theta}=\frac{1}{2}\bar{\theta}^{AB}J_{AB}%
\,,\nonumber\\
F  &  =\frac{1}{2}\Omega^{AB}J_{AB}\,,\ \ \bar{F}=\frac{1}{2}\bar{\Omega}%
^{AB}J_{AB}\,, \label{LT1}%
\end{align}
where $\omega^{AB}$\ and $\bar{\omega}^{AB}$ are two Lorentz spin connections
and $\theta^{AB}=\omega^{AB}-\bar{\omega}^{AB}$. 
The topological terms can be written as,
\begin{align}
\left\langle F^{p}\right\rangle  &  =\frac{1}{2^{p}}\varepsilon_{A_{1}\ldots
A_{2p}}\Omega^{A_{1}A_{2}}\cdots\Omega^{A_{2p-1}A_{2p}}=\frac{1}{2^{p}%
}\varepsilon_{2p}\left(  \Omega\right)  \,,\nonumber\\
\left\langle \bar{F}^{p}\right\rangle  &  =\frac{1}{2^{p}}\varepsilon
_{A_{1}\ldots A_{2p}}\bar{\Omega}^{A_{1}A_{2}}\cdots\bar{\Omega}%
^{A_{2p-1}A_{2p}}=\frac{1}{2^{p}}\varepsilon_{2p}\left(  \bar{\Omega}\right)
\,, \label{LT4}%
\end{align}
where $\varepsilon_{A_{1}\ldots A_{2p}}=\left\langle J_{A_{1}A_{2}}\cdots
J_{A_{2p-1}A_{2p}}\right\rangle $ and where $\varepsilon_{2p}\left(
\Omega\right)  \,$, $\varepsilon_{2p}\left(  \bar{\Omega}\right)  $ are called
Euler topological terms. Using $\bar{\theta}^{2}=\frac{1}{2}\bar{\theta
}_{\ C}^{A}\bar{\theta}^{CB}J_{AB}$, which can be shown using (\ref{L_com}),
the interpolating strength field $F_{t}=\frac{1}{2}\Omega_{\left(  t\right)
}^{AB}J_{AB}$ can be written in the following alternative forms,%
\begin{align}
\Omega_{\left(  t\right)  }^{AB}  &  =\bar{\Omega}^{AB}+t\bar{D}\bar{\theta
}^{AB}+t^{2}\bar{\theta}_{\ C}^{A}\bar{\theta}^{CB}\label{Rt_1}\\
&  =\Omega^{AB}+\left(  t-1\right)  D\bar{\theta}^{AB}+\left(  t-1\right)
^{2}\bar{\theta}_{\ C}^{A}\bar{\theta}^{CB}\label{Rt_2}\\
&  =t\Omega^{AB}+\left(  1-t\right)  \bar{\Omega}^{AB}-t\left(  1-t\right)
\bar{\theta}_{\ C}^{A}\bar{\theta}^{CB} \label{Rt_3}\,.%
\end{align}
Thus, the transgression form is given by $T^{\left(  2p-1\right)  }\left(
\omega,\bar{\omega}\right)  =\frac{1}{2^{p}}\mathcal{T}^{\left(  2p-1\right)
}\left(  \omega,\bar{\omega}\right)$, where%
\begin{equation}
\mathcal{T}^{\left(  2p-1\right)  }\left(  \omega,\bar{\omega}\right)
=p\int_{0}^{1}dt\varepsilon_{A_{1}\ldots A_{2p}}\bar{\theta}^{A_{1}A_{2}%
}\Omega_{\left(  t\right)  }^{A_{3}A_{4}}\cdots\,\Omega_{\left(  t\right)
}^{A_{2p-1}A_{2p}}\,, \label{LT5}%
\end{equation}
so the Chern-Weyl theorem for the Lorentz symmetry \ reads,%
\begin{equation}
\varepsilon_{2p}\left(  \Omega\right)  -\varepsilon_{2p}\left(  \bar{\Omega
}\right)  =d\mathcal{T}^{\left(  2p-1\right)  }\left(  \omega,\bar{\omega
}\right)  \,. \label{LT6}%
\end{equation}

\section{Tensorial version of the CW theorem in $D=2$\label{Conf_spaces}}

In order to formulate the two-dimensional Chern-Weil theorem in tensorial
language, one can use the fact that all 2-dimensional pseudo-Riemannian
manifolds are conformally related. To do this, let us consider a given metric
$g_{\mu\nu}^{\left(  0\right)  }$ and two functions $u\left(  x^{\mu}\right)
$ and $\bar{u}\left(  x^{\mu}\right)  $ such that the metrics of the manifolds
$\mathcal{M}_{2}$ and $\mathcal{\bar{M}}_{2}$ are given by%
\begin{equation}
g_{\mu\nu}=e^{2u}g_{\mu\nu}^{\left(  0\right)  }\ ,\ \ \ \bar{g}_{\mu\nu
}=e^{2\bar{u}}g_{\mu\nu}^{\left(  0\right)  }\,. \label{conf_1}%
\end{equation}
A direct calculation of the Ricci tensors leads the following relation (in
$D=2$)%
\begin{equation}
R_{\mu\nu}^{\left(  0\right)  }=R_{\mu\nu}+g_{\mu\nu}\square u=\bar{R}_{\mu
\nu}+\bar{g}_{\mu\nu}\bar{\square}\bar{u}\,. \label{conf_2}%
\end{equation}
Using $g^{\left(  0\right)  \mu\nu}=e^{2u}g^{\mu\nu}=e^{2\bar{u}}\bar{g}%
^{\mu\nu}$ we obtain the following relation between the Ricci scalars,%
\begin{equation}
R^{\left(  0\right)  }=e^{2u}\left(  R+2\square u\right)  =e^{2\bar{u}}\left(
\bar{R}+2\bar{\square}\bar{u}\right)  \,. \label{conf_3}%
\end{equation}
Thus, with the usual properties of the operators $\square$ and $\bar{\square}$
the last relation can be equivalently written as,%
\begin{align}
R  &  =-2\square u+e^{-2u}R^{\left(  0\right)  }=-\frac{2}{\sqrt{-g}}%
\partial_{\mu}\left(  \sqrt{-g}g^{\mu\nu}\partial_{\nu}u\right)
+e^{-2u}R^{\left(  0\right)  }\,,\nonumber\\
\bar{R}  &  =-2\bar{\square}\bar{u}+e^{-2\bar{u}}R^{\left(  0\right)  }%
=-\frac{2}{\sqrt{-\bar{g}}}\partial_{\mu}\left(  \sqrt{-\bar{g}}\bar{g}%
^{\mu\nu}\partial_{\nu}\bar{u}\right)  +e^{-2\bar{u}}R^{\left(  0\right)  }\,.
\label{conf_4}%
\end{align}
Then, the difference of the topological terms $\sqrt{-g}Rd^{2}x$ and
$\sqrt{-\bar{g}}\bar{R}d^{2}x$ is given by%
\begin{equation}
\left(  \sqrt{-g}R-\sqrt{-\bar{g}}\bar{R}\right)  d^{2}x=\left[
-2\partial_{\mu}\left(  \sqrt{-g}g^{\mu\nu}\partial_{\nu}u-\sqrt{-\bar{g}}%
\bar{g}^{\mu\nu}\partial_{\nu}\bar{u}\right)  +\sqrt{-g}e^{-2u}R^{\left(
0\right)  }-\sqrt{-\bar{g}}e^{-2\bar{u}}R^{\left(  0\right)  }\right]  d^{2}x\,.
\label{conf_5}%
\end{equation}
Finally, using the relations
\begin{align}
\sqrt{-g}  &  =e^{2u}\sqrt{-g^{\left(  0\right)  }}\ ,\ \ \ \sqrt{-\bar{g}%
}=e^{2\bar{u}}\sqrt{-g^{\left(  0\right)  }} \,, \nonumber\\
g_{\mu\nu}  &  =e^{2U}\bar{g}_{\mu\nu}\,,\ \ \ \bar{g}^{\mu\nu}=e^{2U}%
g^{\mu\nu}\,,\ \ \ \sqrt{-\bar{g}}=e^{-2U}\sqrt{-g}\,,\ \ \ U=u-\bar{u}\,,
\label{conf_6}%
\end{align}
that can be easily derived from (\ref{conf_1}), we get%
\begin{equation}
\left(  \sqrt{-g}R-\sqrt{-\bar{g}}\bar{R}\right)  d^{2}x=-\partial_{\mu}%
v^{\mu}d^{2p}x\,, \label{conf_7}%
\end{equation}
where
\begin{equation}
v^{\mu}=2\sqrt{-g}g^{\mu\nu}\partial_{\nu}U\,. \label{conf_8}%
\end{equation}

Eq. (\ref{conf_7}), with $v^{\mu}$ given by (\ref{conf_8}), represents a
tensorial version of the Chern-Weil theorem which is free of objects coming
from the vielbein formalism and that depends only on the metrics $g_{\mu\nu}$,
$\bar{g}_{\mu\nu}$ and the conformal factor $U$ relating the metrics of
$\mathcal{M}_{2}$ and $\mathcal{\bar{M}}_{2}\,$. Although this is an
interesting result, which is valid for any given pair of manifold
$\mathcal{M}_{2}$ and $\mathcal{\bar{M}}_{2}$, a generalization to higher
dimensions is not possible because the fact that all metrics are conformally
equivalent is an accident that happens only in $D=2$.

A generalization to $D=2p$ might work only under the assumption that
$\mathcal{M}_{2}$ and $\mathcal{\bar{M}}_{2}$ are conformally equivalent. We
leave that problem for a possible future work.

\section{Relation between vielbeins of different spaces\label{App_Rel_vielb}}

The orthotormal inverse vielbeins $e_{A}^{\mu}$ at $P\in\mathcal{M}_{D}$ and
$\bar{e}_{A}^{\mu}$ at $\bar{P}\in\mathcal{\bar{M}}_{D}$, can be defined by
means of two different coordinate transformations, one in $P$ the other at
$\bar{P}\,$,
\begin{equation}
x^{\mu}=x^{\mu}\left(  y^{A}\right)  \ \ \text{and\ \ }x^{\mu}=x^{\mu}\left(
\bar{y}^{A}\right)  \,, \label{ct}%
\end{equation}
so that the metrics in $\mathcal{M}_{D}$ and $\mathcal{\bar{M}}_{D}$ become
Minkowski at $P$ and $\bar{P}$,%
\begin{align}
e_{A}^{\mu}e_{B}^{\nu}g_{\mu\nu}  &  =\eta_{AB}\,,\ \text{with }e_{A}^{\mu
}\left(  P\right)  =\dfrac{\partial x^{\mu}}{\partial y^{A}}\left(  P\right)
\,,\label{vielb1}\\
\bar{e}_{A}^{\mu}\bar{e}_{B}^{\nu}\bar{g}_{\mu\nu}  &  =\eta_{AB}%
\,,\ \text{with}\ \bar{e}_{A}^{\mu}\left(  \bar{P}\right)  =\dfrac{\partial
x^{\mu}}{\partial\bar{y}^{A}}\left(  \bar{P}\right)  \,. \label{vielb2}%
\end{align}
The coordinates $y^{A}$ are $\bar{y}^{A}$ must be different, otherwise we are
lead to the contradiction $g_{\mu\nu}=\bar{g}_{\mu\nu}\,$. However, the
existence of the maping $\sigma$ introduced in Section \ref{S_L_vs_S}\ implies
that $y^{A}$ and $\bar{y}^{A}$ are smoothly related. Indeed, using that the
relations (\ref{ct}) are invertible, we can write%
\begin{equation}
y^{A}=y^{A}\left(  x^{\mu}\right)  =y^{A}\left(  x^{\mu}\left(  \bar{y}%
^{B}\right)  \right)  \equiv y^{A}\left(  x^{\mu},\bar{y}^{B}\right)  \,.
\label{noLorentz_ct}%
\end{equation}

In the last expression we wrote a explicitly a dependence on the coordinate
$x^{\mu}$. The reason is that the relations (\ref{ct}) are not simple
coordinate transformations made in a given patch of each manifold. They are
rather one pair of coordinate transformation for each couple of points
$\left(  P,\bar{P}\right)  $. As $y^{A}$ and $\bar{y}^{A}$ are cartesian
coordinates of two Minkowski spaces tangent to the points $P$ and $\bar{P}$
(i.e., such that the metric in both cases is $\eta_{AB}$) we see that $y^{A}$
must be a linear, point dependent function of $\bar{y}^{A}\,$. Without loss of
generality we can assume that is also homogeneous, namely $y^{A}=K_{\ B}^{A}\left(  x\right)  \bar{y}^{B}$, so the vielbeins are related by%
\begin{equation}
e^{A}=K_{\ B}^{A}\left(  x\right)  \bar{e}^{B}\,. \label{noL_vielbT}%
\end{equation}
Thus, for any given pair of manifolds $\mathcal{M}_{D}$ and $\mathcal{\bar{M}%
}_{D}\,$, the matrix $K$ can be directly solved from (\ref{noL_vielbT}) as
\[
K_{\ B}^{A}\left(  x\right)  =e_{\mu}^{A}\left(  x\right)  \bar{e}_{B}^{\mu
}\left(  x\right)  \,.
\]
Clearly, the matrix $K_{\ B}^{A}\left(  x\right)  $\ cannot be a Lorentz
rotation, otherwise $ds^{2}$ and $d\bar{s}^{2}$ would coincide. To understand
better this result, we remark that in the coordinates $y^{A}\,$, the metric of
the Minkowski tangent space $T_{P}\left(  \mathcal{M}_{D}\right)  $ in $P$ is
given by $\eta_{AB}$. If we use the coordinates $\bar{y}^{A}=\left(
K^{-1}\right)  _{\ B}^{A}\left(  x\right)  y^{B}$, then the metric of
$T_{P}\left(  \mathcal{M}_{D}\right)  $ is given by%
\begin{equation}
g_{AB}\left(  x\right)  =K_{\ A}^{C}\left(  x\right)  K_{\ B}^{D}\left(
x\right)  \eta_{CD}\neq\eta_{AB}\,, \label{g_NL_1}%
\end{equation}
i.e., the Lorentz metric $\eta_{AB}$ is not preserved because the considered
coordinate transformation is not of the Lorentz type. The same applies for the
tangent space $T_{\bar{P}}\left(  \mathcal{\bar{M}}_{D}\right)  $, whose
metric in the coordinates $\bar{y}^{A}$ is given by $\eta_{AB}$ while in
coordinates $y^{A}=K_{\ B}^{A}\left(  x\right)  \bar{y}^{B}$ it is given by,%
\begin{equation}
\bar{g}_{AB}\left(  x\right)  =\left(  K^{-1}\right)  _{\ A}^{C}\left(
x\right)  \left(  K^{-1}\right)  _{\ B}^{D}\left(  x\right)  \eta_{CD}\neq
\eta_{AB}\,. \label{g_NL_2}%
\end{equation}

As an example, for the case of static spherically symmetric manifolds with
metrics,%
\begin{equation}
ds^{2}=-f^{2}\left(  r\right)  dt^{2}+\frac{1}{h^{2}\left(  r\right)  }%
dr^{2}+r^{2}d\Omega_{D-2}^{2}\,,\ \ \ d\bar{s}^{2}=-\bar{f}^{2}\left(
r\right)  dt^{2}+\frac{1}{\bar{h}^{2}\left(  r\right)  }dr^{2}+r^{2}%
d\Omega_{D-2}^{2}\,, \label{SSS_bar}%
\end{equation}
we obtain,%
\[
K=\left(  K_{\ B}^{A}\right)  =\left(
\begin{array}
[c]{ccc}%
f\,/\,\bar{f} & 0 & 0\\
0 & \bar{h}\,/\,h & 0\\
0 & 0 & \delta_{\ b}^{a}%
\end{array}
\right)  \,,
\]
with $a,b=2,\ldots,D-1\,$. Thus, we see explicitly that $K^{T}\eta K\neq\eta$
unless $f=\bar{f}$ and $h=\bar{h}$.

\section{Geometric properties of $\tilde{\omega}_{\ B}^{A}$%
\label{Geom_prop_hyb}}

Let us associate $\tilde{\omega}_{\ B}^{A}$ with an auxiliary manifold
$\mathcal{\tilde{M}}$ with metric $\tilde{g}_{\mu\nu}$ and affine connection
$\tilde{\Gamma}_{\mu\gamma}^{\alpha}$ so that the vielbein $\tilde{e}^{A}$ and
the hybrid spin connection $\tilde{\omega}_{\ B}^{A}$ verify the usual
relations\footnote{We also choose a smooth mapping allowing us to use the same
coordinates $x^{\mu}$ for each point $P\in\mathcal{M}_{D}$ and $\tilde{P}%
\in\mathcal{\tilde{M}}_{D}\,$.}%
\begin{align}
\tilde{g}_{\mu\nu}  &  =\tilde{e}_{\mu}^{A}\tilde{e}_{\nu}^{B}\eta
_{AB}\,,\label{hyb3}\\
\tilde{\omega}_{\mu B}^{A}  &  =\tilde{e}_{\alpha}^{A}\tilde{e}_{B}^{\gamma
}\tilde{\Gamma}_{\mu\gamma}^{\alpha}+\tilde{e}_{\alpha}^{A}\partial_{\mu
}\tilde{e}_{B}^{\alpha}\,, \label{hyb4}%
\end{align}
and allow to define the curvature and torsion two-forms as
\begin{equation}
\tilde{\Omega}_{\ B}^{A}\equiv d\tilde{\omega}_{\ B}^{A}+\tilde{\omega}%
_{\ C}^{A}\tilde{\omega}_{\ B}^{C}\,,\ \ \ \tilde{T}^{A}\equiv\tilde{D}%
\tilde{e}^{A}\,. \label{C_T_hyb}%
\end{equation}
A direct consequence of the definition (\ref{hyb1}) is that $\mathcal{\tilde
{M}}$ is not a metric compatible manifold, i.e., $\tilde{\nabla}\tilde{g}%
_{\mu\nu}\neq0$ and thus $\tilde{\Gamma}_{\mu\gamma}^{\alpha}$ is not the
Christoffel symbol. To see this, we first notice that $\tilde{\omega}^{AB}$ is
not antisymmetric (or, equivalently $\tilde{D}\eta_{AB}\neq0$). Indeed, Eq.
(\ref{hyb1}) can be written as $\tilde{\omega}_{\mu}^{AB}=e_{\alpha}^{A}%
\bar{\nabla}_{\mu}e^{B\alpha}=-e^{B\alpha}\bar{\nabla}_{\mu}e_{\alpha}^{A}$
and using $\bar{\nabla}_{\lambda}g_{\mu\nu}\neq0$ (with $\bar{\nabla}%
_{\lambda}$ being the covariant derivative with respect to $\bar{\Gamma}%
_{\mu\nu}^{\alpha}\,$) one gets\footnote{The fact that $\tilde{\omega}^{AB}$
is not antisymmetric holds even in the case where $\mathcal{M}_{D}$ and
$\mathcal{\bar{M}}_{D}$ are static spherically symmetric spacetimes with
metrics (\ref{SSS_bar}). In that case Eq. (\ref{hyb1}) leads, for example,
$\tilde{\omega}_{\ 0}^{0}=-\tilde{\omega}^{00}=\left(  \bar{f}^{\prime}%
/\bar{f}-f^{\prime}/f\right)  dr$ which does not vanish unless $\mathcal{M}%
_{D}$ and $\mathcal{\bar{M}}_{D}$ coincide.}%
\begin{equation}
\tilde{\omega}_{\mu}^{AB}=-e^{B\alpha}\bar{\nabla}_{\mu}e_{\alpha}^{A}%
\neq-e_{\alpha}^{B}\bar{\nabla}_{\mu}e^{A\alpha}=-\tilde{\omega}_{\mu}^{BA}\,.
\label{hyb4c}%
\end{equation}
Similarly, Eq. (\ref{hyb4}) can be rewritten as $\tilde{\omega}_{\mu}%
^{AB}=\tilde{e}_{\alpha}^{A}\tilde{\nabla}_{\mu}\tilde{e}^{B\alpha}=-\tilde
{e}^{B\alpha}\tilde{\nabla}_{\mu}\tilde{e}_{\alpha}^{A}$ and then, using
$\tilde{\omega}_{\mu}^{AB}\neq-\tilde{\omega}_{\mu}^{BA}$ we have%
\[
\tilde{e}^{B\alpha}\tilde{\nabla}_{\mu}\tilde{e}_{\alpha}^{A}\neq\tilde
{e}_{\alpha}^{B}\tilde{\nabla}_{\mu}\tilde{e}^{A\alpha}\,,%
\]
which clearly implies $\tilde{\nabla}\tilde{g}_{\mu\nu}\neq0$.

The non metricity of $\mathcal{\tilde{M}}$ does not necessarily represent a
problem, because it can be thought just as an auxiliary manifold allowing to
translate Lorentz invariant expressions constructed with $\tilde{\theta}^{AB}%
$, to tensorial language. Then, is necessary to show that $\mathcal{\tilde{M}%
}$ carries no new independent information, i.e., that its geometry can be
completely fixed in terms of geometrical quantities of $\mathcal{M}$ and
$\mathcal{\bar{M}}$. Indeed, from the definition (\ref{hyb1}) the components
$\tilde{\omega}_{\mu B}^{A}$ can be completely solved in terms of geometrical
quantities of $\mathcal{M}$ and $\mathcal{\bar{M}}$ and plugging this in
(\ref{hyb4}) together with the torsionless condition, $\tilde{\Gamma}%
_{\mu\gamma}^{\alpha}=\tilde{\Gamma}_{\gamma\mu}^{\alpha}$, one can solve the
independent components of $\tilde{e}_{\alpha}^{A}$ and $\tilde{\Gamma}%
_{\mu\gamma}^{\alpha}$. Even if this is a hard task, due to the big number
of unknown functions that must be solved when $\tilde{\Gamma}_{\mu\gamma
}^{\alpha}$ is not the Christoffel symbol, it can always be done\footnote{Due
to the invariance under Lorentz rotations, there are only $D\left(
D+1\right)  /2$ independent components in the vielbein $\tilde{e}_{\alpha}%
^{A}$, which are the same that characterize the metric $g_{\mu\nu}$. For the
affine connection $\tilde{\Gamma}_{\mu\gamma}^{\alpha}$, which is not the
Christoffel symbol, the tosionless condition implies that there may be up to
$D^{2}\left(  D+1\right)  /2$ independent components more. They give a total
of $D\left(  D+1\right)  ^{2}/2$ independent functions which can always be
solved with the $D^{3}$ independent equations (\ref{hyb4}).}.

The real problem about the hybrid connection (\ref{hyb1}) is related with the
definition of the curvature made in (\ref{C_T_hyb}). Even if the Bianchi
identities $\tilde{D}\tilde{\Omega}_{\ B}^{A}=0$ and $\tilde{D}\tilde{T}%
^{A}=\tilde{\Omega}_{\ B}^{A}\tilde{e}^{B}$ are satisfied, for%
\begin{equation}
\tilde{\Omega}^{AB}\equiv\eta^{BC}\tilde{\Omega}_{\ C}^{A}=d\tilde{\omega
}^{AB}+\tilde{\omega}_{\ C}^{A}\tilde{\omega}^{CB} \label{hyb5}%
\end{equation}
we have instead%
\begin{equation}
\tilde{D}\tilde{\Omega}^{AB}=d\tilde{\Omega}^{AB}+\tilde{\omega}_{\ C}%
^{A}\tilde{\Omega}^{CB}+\tilde{\omega}_{\ C}^{B}\tilde{\Omega}^{AC}=\left(
\tilde{D}\eta^{BC}\right)  \tilde{\Omega}_{\ C}^{A}\neq0\,. \label{hyb5b}%
\end{equation}
The reason is that the curvature $\tilde{\Omega}^{AB}$, as given by Eq.
(\ref{hyb5}), does not comes from the definition of a strenght field in the
fiber bundle formulation of gauge theories (a brief review was given in Appendix \ref{App_Tran}). Briefly, this means that if the hybrid gauge
connection one-form is defined as $\tilde{A}=\frac{1}{2}\tilde{\omega}%
^{AB}J_{AB}$, where $J_{AB}$ are the antisymmetric Lorentz generators
satisfying the commutation relations (\ref{L_com}), then its strenght field
must be defined as
\begin{equation}
\tilde{F}\equiv d\tilde{A}+\tilde{A}\tilde{A}=\frac{1}{2}\tilde{\Omega}%
^{AB}J_{AB}\,, \label{hyb6}%
\end{equation}
which leads%
\begin{equation}
\tilde{\Omega}^{AB}=d\tilde{\omega}^{\left[  AB\right]  }+\eta_{CD}%
\tilde{\omega}^{\left[  AC\right]  }\tilde{\omega}^{\left[  DB\right]
}\,,\text{ \ \ with \ }\tilde{\omega}^{\left[  AB\right]  }=\frac{1}{2}\left(
\tilde{\omega}^{AB}-\tilde{\omega}^{BA}\right)\,,  \label{hyb7}%
\end{equation}
which is clearly different from (\ref{hyb5}). Defined by (\ref{hyb7}),
$\tilde{\Omega}^{AB}$ is antisymmetric by construction and it should be
regarded as the strenght field for $\tilde{\omega}^{\left[  AB\right]  }$
rather than $\tilde{\omega}^{AB}$. Indeed, it is direct to show that under a
Lorentz transformation%
\[
\tilde{\omega}_{\mu}^{[A^{\prime}B^{\prime}]}=\Lambda_{\ C}^{A^{\prime}%
}\Lambda_{\ D}^{B^{\prime}}\tilde{\omega}_{\mu}^{\left[  CD\right]  }%
+\eta^{CD}\Lambda_{\ C}^{A^{\prime}}\partial_{\mu}\Lambda_{\ D}^{B^{\prime}%
}\,,
\]
and thus, $\tilde{\omega}^{\left[  AB\right]  }$ is also a well-defined
Lorentz connection. In addition, the Bianchi identity comming from the gauge
formulation is then given by $\mathrm{\tilde{D}}\tilde{F}\equiv d\tilde
{F}+[\tilde{A},\tilde{F}]=0$ (where $\mathrm{\tilde{D}}$ denotes the covariant
derivative with respect to the gauge field $\tilde{A}\,$), and leads%
\begin{equation}
\tilde{D}\tilde{\Omega}^{AB}=d\tilde{\Omega}^{AB}+\eta_{CD}\tilde{\omega
}^{\left[  AC\right]  }\tilde{\Omega}^{\left[  DB\right]  }+\eta_{CD}%
\tilde{\omega}^{\left[  BC\right]  }\tilde{\Omega}^{\left[  AD\right]  }=0\,.
\label{hyb7b}%
\end{equation}

To avoid antisymmetrization brackets, we can define the \textit{antisymmetric
hybrid} spin connnection as\footnote{Notice that, just as it happens in Eq.
(\ref{hyb2}) for $\tilde{\theta}^{AB}$, if we define $\breve{\theta}%
^{AB}\equiv\omega^{AB}-\breve{\omega}^{AB}$ we still can factorize two
vielbeins $\breve{\theta}^{AB}=e_{\alpha}^{\left[  A\right.  }e_{\beta
}^{\left.  B\right]  }\Delta_{\mu}^{\alpha\beta}dx^{\mu}$ and the
antisymmetrization bracket changes anything, because this kind of object is
usually multiplied by the antisymmetric tensor $\varepsilon_{A_{1}\ldots
A_{D}}$ in a Lorentz invariant expression.
\par
For consistency, we also notice that the antisymmetric property of
$\breve{\omega}^{AB}$ holds independent from the fact that $\bar{\nabla}_{\mu
}g_{\alpha\beta}\neq0$. Indeed, those kind of terms that make $\tilde{\omega
}^{AB}$ not been antisymmetric, now are cancelled%
\[
\breve{\omega}_{\mu}^{AB}=\frac{1}{2}\left(  e_{\alpha}^{A}\bar{\nabla}_{\mu
}e^{B\alpha}-e_{\alpha}^{B}\bar{\nabla}_{\mu}e^{A\alpha}\right)  =\frac{1}%
{2}\left(  e^{A\alpha}e^{B\beta}\bar{\nabla}_{\mu}\left(  g_{\alpha\beta
}-g_{\beta\alpha}\right)  -\left(  e_{\alpha}^{B}\bar{\nabla}_{\mu}e^{A\alpha
}-e_{\alpha}^{A}\bar{\nabla}_{\mu}e^{B\alpha}\right)  \right)  =-\breve
{\omega}_{\mu}^{BA}\,.%
\]
}%
\begin{equation}
\breve{\omega}_{\mu}^{AB}\equiv\tilde{\omega}_{\mu}^{\left[  AB\right]
}=\frac{1}{2}\left(  e_{\alpha}^{A}e^{B\gamma}\bar{\Gamma}_{\mu\gamma}%
^{\alpha}+e_{\alpha}^{A}\partial_{\mu}e^{B\alpha}-e_{\alpha}^{B}e^{A\gamma
}\bar{\Gamma}_{\mu\gamma}^{\alpha}-e_{\alpha}^{B}\partial_{\mu}e^{A\alpha
}\right)  \,. \label{hyb8}%
\end{equation}
The strenght field $\breve{F}=d\breve{A}+\breve{A}\breve{A}=\frac{1}{2}%
\breve{R}^{AB}J_{AB}$ associated with the gauge connection $\breve{A}=\frac
{1}{2}\breve{\omega}^{AB}J_{AB}\,$ lead the following definition for Lorentz
curvature two-form $\breve{\Omega}^{AB}\equiv d\breve{\omega}^{AB}%
+\breve{\omega}_{\ C}^{A}\breve{\omega}^{CB}\,$, with $\breve{\omega}%
_{\ B}^{A}=\eta_{BC}\breve{\omega}^{AC}$, while the Bianchi identity
$\mathrm{\check{D}}\breve{F}\equiv d\breve{F}+[\breve{A},\breve{F}]=0\,$
reads $\breve{D}\breve{\Omega}^{AB}=d\breve{\Omega}^{AB}+\breve{\omega}%
_{\ C}^{A}\breve{\Omega}^{CB}+\breve{\omega}_{\ C}^{B}\breve{\Omega}^{AC}%
=0\,$. Using $\breve{\Omega}_{\ B}^{A}\equiv\eta_{BC}\breve{\Omega}^{AC}$
leads the usual expression for the Lorentz curvature%
\begin{equation}
\breve{\Omega}_{\ B}^{A}=d\breve{\omega}_{\ B}^{A}+\breve{\omega}_{\ C}%
^{A}\breve{\omega}_{\ B}^{C}\,, \label{hyb9}%
\end{equation}
and then the Bianchi identity can also be written as $\breve{D}\breve{\Omega
}_{\ B}^{A}=0$, because $\breve{D}\eta_{AB}=0$ due to the fact that
$\breve{\omega}$ is antisymmetric by construction. 

Due to the antisymmetry of the Lorentz generators $J_{AB}$ one sees that $\breve{A}=\tilde{A}$ and
$\breve{F}=\tilde{F}$. Thus, the introduction of (\ref{hyb8}) can be thought
just as a change of notation that allows to write the expressions (\ref{hyb9})
which is free of antisymmetrization brackets. However, in Section \ref{S_A_tens_version} this notation has proved to be very useful to determine the geometrical properties of the auxiliary pseudo-Riemannian manifold $\mathcal{\breve{M}}$ that allowed us to give a tensorial formulation of the CW theorem.
%in what follows we will see that it also useful to fully characterize the geometrical properties of the auxiliary manifold that will be used to give a tensorial formulation of the CW theorem.

\end{document}